\documentclass[12pt,letterpaper]{JHEP3}

\usepackage{amscd,amsmath,amssymb,amsfonts,xspace,mathrsfs}
\usepackage{epsfig}

\numberwithin{equation}{section}

\def\varpi{t}

\def\sign{{\rm sign}}

\def\Im{\,{\rm Im}\,}
\def\Re{\,{\rm Re}\,}

\def\rangl{\right\rangle   }
\def\langl{\left\langle  }
\def\({\left(}
\def\){\right)}
\def\[{\left[}
\def\]{\right]}
\def\hf{{1\over 2}}

\def\<{\left\langle}
\def\>{\right\rangle}

\renewcommand{\d}{\mathrm{d}}
\newcommand{\de}{\mathrm{d}}

\newcommand{\I}{\mathrm{i}}

\newcommand{\p}{\partial}

\newcommand{\cV}{\mathcal{V}}
\newcommand{\cC}{\mathcal{C}}
\newcommand{\cS}{\mathcal{S}}
\newcommand{\cG}{\mathcal{G}}
\newcommand{\cK}{\mathcal{K}}
\newcommand{\cM}{\mathcal{M}}
\newcommand{\cW}{\mathcal{W}}
\newcommand{\cN}{\mathcal{N}}

\newcommand{\cX}{\mathcal{X}}

\newcommand{\cR}{\mathcal{R}}
\newcommand{\cT}{\mathcal{T}}
\newcommand{\cJ}{\mathcal{J}}

\newcommand{\cY}{\mathcal{Y}}

\DeclareSymbolFont{AMSa}{U}{msa}{m}{n}
\DeclareSymbolFont{AMSb}{U}{msb}{m}{n}
\DeclareMathSymbol{\fieldR}{\mathalpha}{AMSb}{"52}

\newcommand{\kahler}{{K\"ahler}\xspace}

\newcommand{\qk}{{quaternion-K\"ahler}\xspace}

\newcommand{\cZ}{\mathcal{Z}}

\newcommand{\cU}{\mathcal{U}}
\newcommand{\cA}{\mathcal{A}}
\newcommand{\cB}{\mathcal{B}}
\newcommand{\cQ}{\mathcal{Q}}

\newcommand{\nn}{\nonumber}

\newcommand{\IR}{\mathbb{R}}
\newcommand{\IC}{\mathbb{C}}
\newcommand{\IZ}{\mathbb{Z}}

\newcommand{\tzeta}{\tilde\zeta}

\newcommand{\txi}{\tilde\xi}

\newcommand{\CP}{\IC P^1}

\def\bea{\begin{eqnarray}}
\def\eea{\end{eqnarray}}
\def\be{\begin{equation}}
\def\ee{\end{equation}}
\def\ba{\begin{align}}
\def\ea{\end{align}}
\def\bse{\begin{subequations}}
\def\ese{\end{subequations}}

\def\bX{\bar X}
\def\bF{\bar F}
\def\bY{\bar Y}

\def\bV{ \bar V }

\def\bZ{\bar Z}

\def\ba{\bar a}

\def\bz{\bar z}

\def\hM{\hat M}

\def\tlgam{\tilde\gamma}

\def\ci#1{c^{[#1]}}
\def\cij#1{c^{[#1]}}

\def\txii#1{{\tilde\xi}^{[#1]}}
\def\ai#1{{\alpha}^{[#1]}}

\def\xii#1{\xi_{[#1]}}
\def\alpi#1{\alpha^{[#1]}}

\def\Hij#1{H^{[#1]}}

\def\Xigi{\Xi_{\gamma}}
\def\Xigipr{\Xi_{\gamma'}}

\newcommand{\Li}{{\rm Li}}

\def\Thkl{\Theta_{\gamma}}

\def\Gg{\cG_{\gamma}}

\def\ellg#1{\ell_{#1}}

\def\Zg{Z_{\gamma}}
\def\bZg{\bar Z_{\gamma}}

\def\hng{\Omega_{\gamma}}
\def\Om#1{\Omega_{#1}}
\def\bOm#1{\bar\Omega_{#1}}

\def\tpmn{\varpi^{m,n}_+}
\def\tmmn{\varpi^{m,n}_-}
\def\tpmmn{\varpi^{m,n}_\pm}

\def\Ig{\cJ_{\gamma}}
\def\Ilg{\cJ^{(1)}}
\def\Ilog#1{\cJ^{(1,#1)}}
\def\Irt{\cJ^{(2)}}
\def\Irat#1{\cJ^{(2,#1)}}

\def\Igp{\Ilog{+}_{\gamma}}
\def\Igm{\Ilog{-}_{\gamma}}
\def\Igpm{\Ilog{\pm}_{\gamma}}
\def\Igamp#1{\Ilog{+}_{#1}}

\def\Igg{\Ilg_{\gamma}}
\def\Iggp{\Ilg_{\gamma'}}

\def\Igam#1{\Ilg_{#1}}
\def\rIgam#1{\Irt_{#1}}
\def\rIgamp#1{\Irat{+}_{#1}}
\def\rIgamm#1{\Irat{-}_{#1}}
\def\rIg{\Irt_{\gamma}}
\def\rIgp{\Irat{+}_{\gamma}}
\def\rIgm{\Irat{-}_{\gamma}}
\def\rIgpm{\Irat{\pm}_{\gamma}}

\def\Ingam#1#2{\cJ^{(#1)}_{#2}}
\def\Insgam#1#2#3{\cJ^{(#1,#2)}_{#3}}

\def\XXint#1#2#3{{\setbox0=\hbox{$#1{#2#3}{\int}$}
\vcenter{\hbox{$#2#3$}}\kern-.5\wd0}}

\def\cij#1{c}
\def\ci#1{c}

\def\vv{v}
\def\vvb{{\bf v}}
\def\vl{v}
\def\bvl{\bar \vl}

\def\Min{M}

\def\Minst{\hM}
\def\Uk{U}
\def\Uin{\mathbf{U}}
\def\Ak{A}
\def\cCf{\cC}

\def\CY{\mathfrak{Y}}
\def\CYm{\mathfrak{\hat Y}}
\def\SK{s\cK_c}

\def\Ncl{N^{\rm cl}}

\def\AI{S_3}
\def\BI{S_1}
\def\CI{S_2}

\def\qr{\sigma_\gamma}
\def\qrp{\sigma_{\gamma'}}
\def\qrg#1{\sigma_{#1}}

\title{Hypermultiplet metric and D-instantons}

\author{Sergei Alexandrov and Sibasish Banerjee
\\
{\it Universit\'e Montpellier 2, Laboratoire Charles Coulomb UMR 5221, F-34095,
Montpellier, France}\\

\vspace*{2mm} {\tt e-mail:
\email{salexand@univ-montp2.fr},
\email{sibasishbanerjee@live.in}
}

\vspace*{-3mm}

}

\abstract{
We use the twistorial construction of D-instantons in Calabi-Yau compactifications of type II string theory
to compute an explicit expression for the metric on the hypermultiplet moduli space affected by these non-perturbative corrections.
In this way we obtain an exact \qk metric which is a non-trivial deformation of the local c-map.
In the four-dimensional case corresponding to the universal hypermultiplet, our metric fits the Tod ansatz
and provides an exact solution of the continuous Toda equation.
We also analyze the fate of the curvature singularity of the perturbative metric by deriving an S-duality invariant equation
which determines the singularity hypersurface after inclusion of the D(-1)-instanton effects.
}

\begin{document}

%-----------------------------------------------------------------------------------
\section{Introduction}

One of the outstanding problems in string theory is to find the effective low energy dynamics for various
classes of compactifications. Whereas from the phenomenological point of view we are primarily interested
in compactifications preserving not more than four supercharges in four dimensions, our understanding of these cases remains still
rudimentary and limited to the weak coupling regime. On the other hand, quantum effects in general, and
non-perturbative effects in particular, which affect the effective action at strong coupling, are known to play an
extremely important role. For instance, one has to take them into account to stabilize all moduli and to get a viable
cosmological models \cite{Kachru:2003aw}, they provide resolution of unphysical singularities in the moduli space \cite{Ooguri:1996me},
and they appear to be a crucial ingredient ensuring various stringy dualities \cite{Witten:1995ex,Green:1997di}.
Therefore, having control over such effects would definitely produce a great impact on different research directions.

In recent years a significant progress has been achieved in understanding of the non-perturbative effective action
resulting from type II string theory compactified on a Calabi-Yau threefold $\CY$. In this case the low energy theory has $N=2$ supersymmetry
and the effective action is completely determined by the geometry of the vector and hypermultiplet (HM) moduli spaces \cite{Bagger:1983tt,deWit:1984px}.
The former is a special \kahler (SK) manifold and is classically exact (no corrections in the string coupling $g_s$), whereas the latter
is \qk (QK) and receives perturbative and non-perturbative $g_s$-corrections.
The progress mentioned above was related with the developments of twistorial methods which provide an efficient parametrization
of QK geometries \cite{MR1327157,Alexandrov:2008ds,Alexandrov:2008nk}. Combining these methods with the symmetries expected to survive at quantum level,
a large class of instanton corrections to the HM moduli space has been found
\cite{RoblesLlana:2006is,RoblesLlana:2007ae,Alexandrov:2008gh,Alexandrov:2009zh,Alexandrov:2010ca,Alexandrov:2012au,Alexandrov:2014mfa,Alexandrov:2014rca}
(see \cite{Alexandrov:2011va,Alexandrov:2013yva} for reviews).
Although the description of few types of instantons remains still unknown, the complete non-perturbative picture
for this class of compactifications seems to be already not far from our reach.

However, these results encode the HM metric in a very inexplicit way: they are formulated in terms of ceratin
holomorphic data on the twistor space, a canonical $\CP$ bundle over the original QK manifold.
In principle, these data contain all geometric information, and the procedure to extract the metric from them is well known.
But it is often quite difficult to realize it in practice.
As a result, an explicit expression for the metric was not known beyond the perturbative approximation.
On the other hand, it might be interesting not only from the pure mathematical point of view, as an example of an {\it exact}
and non-trivial QK metric, but also from the perspective of physical applications such as moduli stabilization
and producing an inflationary potential by gauging some of the isometries on the moduli space \cite{Polchinski:1995sm,deWit:2001bk}.

In this paper we fill this gap by computing the HM moduli space metric in the presence of D-instanton corrections.
More precisely, our result applies in the two cases. First, if one includes only electrically charged D-instantons
(in the type IIA formulation these are instantons coming from D2-branes wrapping A-cycles in $H_3(\CY,\IZ)$,
whereas in type IIB they correspond to D(-1) and D1-instantons), the obtained metric is valid to all orders in the instanton expansion
and thus it is an {\it exact} \qk metric. In the second case, one can consider all D-instantons, but then the resulting metric
is valid only in the one-instanton approximation.

In the special case of one hypermultiplet, known as {\it universal hypermultiplet}, our results describe a four-dimensional QK manifold
with one continuous isometry. By a proper choice of coordinates the metric on such spaces can always be put
in the so-called Tod ansatz, which is described by one real function satisfying the non-linear Toda differential equation \cite{MR1423177}.
We show that the D-instanton corrected HM metric, which we computed, perfectly fits this ansatz and the potential, one derives from it,
does solve the Toda equation. This provides a very non-trivial consistency check of our results.

Besides, we reconsider the issue of a curvature singularity, which was studied before
in \cite{Alexandrov:2009qq}. The singularity appears after inclusion of the one-loop $g_s$-correction in the tree level HM metric
\cite{Antoniadis:2003sw,Robles-Llana:2006ez,Alexandrov:2007ec}. Knowing the metric in the presence of D-instantons, we are able to
study how they affect this singularity. In particular, we derive an equation for its position in the moduli space.
Furthermore, restricting to the D(-1)-instantons on the type IIB side, we rewrite this equation in the form which is explicitly invariant
under the $SL(2,\IZ)$ duality group. This allows us to relate the weak and strong coupling regions and establish that the singularity
is still present. This result suggests that perhaps the singularity is resolved only in the full non-perturbative metric
which includes not only D-instantons, but also corrections from NS5-branes wrapping the whole Calabi-Yau \cite{Becker:1995kb}.

The paper is organized as follows. In the next section we briefly review some facts about the HM moduli space.
In particular, we explain the twistorial construction of the D-instantons. In section \ref{sec-metric}, starting from this construction,
we compute the explicit expression for the D-instanton corrected metric.
In section \ref{sec-UHM} we specialize this result to the case of the universal hypermultiplet.
Section \ref{sec-sing} is devoted to the analysis of the curvature
singularity. Finally, section \ref{sec-disc} presents our conclusions. In a few appendices we provide some details of the calculations.

%------------------------------------------------------------------------------------
\section{Hypermultiplet moduli space}
\label{sec-HM}

\subsection{Perturbative metric in type IIA}
\label{subsec-pertHM}

The HM moduli space $\cM_H$ is the target space of the non-linear sigma-model describing the dynamics of the scalar fields constituting
the bosonic sector of the hypermultiplets in a theory with $N=2$ supersymmetry. If the supersymmetry is local, i.e. the hypermultiplets are coupled
to $N=2$ supergravity, $\cM_H$ must be a \qk manifold \cite{Bagger:1983tt}, which means that its holonomy group is contained in $Sp(n)\times SU(2)$
where $n$ is the number of hypermultiplets and $\dim_{\IR} \cM_H=4n$. In the case where the theory we are describing emerges from type II string theory
compactified on a Calabi-Yau $\CY$, $\cM_H$ comes equipped with a set of preferable coordinates originating in the geometry of the compactification.

Let us concentrate on the type IIA formulation. Then the low energy effective theory contains $n=h^{2,1}(\CY) +1$ hypermultiplets
and their scalar fields have the following interpretation:
\begin{itemize}
\item
the fields $z^a$ ($a=1,\dots,h^{2,1}$) parametrizing the deformations of the complex structure of $\CY$;

\item
the RR-fields $\zeta^\Lambda,\tzeta_\Lambda$ ($\Lambda=0,\dots,h^{2,1}$)
arising as period integrals of the RR 3-form of type IIA string theory over a symplectic basis of cycles in $H_3(\CY,\IZ)$;

\item
the four-dimensional dilaton $e^\phi = 1/g^2_{(4)}$;

\item
the NS-axion $\sigma$ which is dual to the $B$-field in four dimensions.
\end{itemize}
The subspace parametrized by $z^a$, which we denote $\SK$, carries a natural special \kahler metric \cite{Craps:1997gp} determined
by the holomorphic prepotential $F(X^\Lambda)$,
a homogeneous function of degree two. In terms of this function the \kahler potential of the special \kahler geometry is given by
\be
\cK= -\log K,
\qquad
K  = -2\Im (\bz^\Lambda F_\Lambda(z)),
\ee
where $F_\Lambda=\p_{X^\Lambda}F$ and we defined $z^\Lambda=(1,z^a)$. Note that this subspace and the total space $\cM_H$
carry an action of the symplectic group. In particular, $(X^\Lambda,F_\Lambda)$ and $(\zeta^\Lambda,\tzeta_\Lambda)$
transform in the vector representation, whereas $\phi$ and $\sigma$ are symplectic invariant.

At tree level the metric on $\cM_H$ is obtained by Kaluza-Klein reduction from ten-dimensional supergravity
and turns out to be determined by the prepotential $F$ \cite{Cecotti:1989qn,Ferrara:1989ik}.
It is known as the {\it c-map} metric which gives a canonical
construction of a QK manifold as a bundle over a special \kahler base.
At perturbative level the HM metric receives a one-loop correction only \cite{Robles-Llana:2006ez}.
It is proportional to the Euler class of the Calabi-Yau, $\chi_{\CY} = 2\(h^{1,1}(\CY) - h^{2,1}(\CY) \)$, and thus induces a one-parameter
deformation of the c-map metric \cite{Antoniadis:1997eg,Gunther:1998sc,Antoniadis:2003sw}.
Its explicit expression has been computed in \cite{Alexandrov:2007ec} and reads as
\be
\begin{split}
\de s^2 =&\, \frac{r+2c}{r^2(r+c)}\, \de r^2
-\frac{1}{r} \(N^{\Lambda\Sigma} - \frac{2(r+c)}{rK}\, z^\Lambda \bz^\Sigma\) \(\de \tzeta_\Lambda - F_{\Lambda\Lambda'} \de \zeta^{\Lambda'}\)
\(\de\tzeta_\Sigma -\bF_{\Sigma\Sigma'} \de \zeta^{\Sigma'}\)
\\
&\,
+\frac{r+c}{16 r^2(r+2c)} \(\de\sigma + \tzeta_\Lambda \de \zeta^\Lambda - \zeta^\Lambda \de\tzeta_\Lambda + 4c \cA_K\)^2
+ \frac{4(r+c)}{r}\, \cK_{a\bar{b}} \de z^a \de \bz^b.
\end{split}
\label{1lmetric}
\ee
Here we denoted $r=e^\phi$, $N_{\Lambda\Sigma} = -2\Im F_{\Lambda\Sigma}$, the matrix $N^{\Lambda\Sigma}$ is its inverse,
$c = -\frac{\chi_\CY}{192\pi}$ is the deformation parameter encoding the one-loop correction,
and $\cA_K$ is the so-called K\"ahler connection on $\SK$
\be
\cA_K =  \frac{\I}{2}\(\cK_a\de z^a-\cK_{\bar a}\de \bz^a\).
\label{kalcon}
\ee
Topologically the metric \eqref{1lmetric} describes a bundle with the two-stage fibration structure
\be
\label{doublefib2}
\IR^+_r \ \times\ \(\begin{array}{rc}
S^{1}_\sigma\ \longrightarrow &\cC(r)
 \\
&\downarrow
\\
\cT_{\zeta,\tzeta}\ \longrightarrow &\cJ_c(\CY)
\\
&\downarrow
\\
&  \SK
\end{array}\).
\ee
Here $\cJ_c(\CY)$ is the so-called intermediate Jacobian with the special \kahler base parametrized by complex structure moduli $z^a$ and
with the fiber given by the torus of RR-fields, $\cT_{\zeta,\tzeta}=H^3(\CY,\IR)/H^3(\CY,\IZ)$. In turn, $\cJ_c(\CY)$ appears as the base for
the circle bundle $\cC(r)$ of the NS-axion, with the curvature given by \cite{Alexandrov:2010np}
\be
\de\(\hf\, D\sigma\)=\omega_{\cT}+\frac{\chi_\CY}{24}\,\omega_{\SK},
\label{curvS}
\ee
where $\omega_{\cT}=\de\zeta_\Lambda\wedge \de\zeta^\Lambda$ and $\omega_{\SK}=-\frac{1}{2\pi}\de\cA$ are the \kahler forms on
$\cT$ and $\SK$, respectively.
The second contribution to the curvature is generated by the one-loop correction.
The metric on the circle bundle parametrically depends on the dilaton $r$,
which contributes to the topology just as a common factor.

Note that the deformation induced at one-loop gives rise to three singularities at $r=0$, $r=-c$ and $r=-2c$.
One can show that the first two can be removed by a coordinate transformation \cite{Alexandrov:2009qq}.
On the other hand, the last one is a true singularity as can be checked by computing the quadratic curvature invariant
$R_{\mu\nu\rho\sigma}R^{\mu\nu\rho\sigma}$. Since the physical moduli space cannot have such singularities,
it must be resolved by non-perturbative effects. In section \ref{sec-sing}, we will discuss the effect of D-instantons on this issue.

\subsection{D-instantons and the twistor space}
\label{subsec-Dinst}

Beyond the perturbative approximation described by the metric \eqref{1lmetric}, the HM moduli space is known to receive instanton corrections
coming from branes wrapping non-trivial cycles of the Calabi-Yau. There are two classes of such corrections corresponding to
the two types of branes in string theory: D-branes and NS5-branes. The former are by now well understood, at least in the type IIA formulation.
The latter have been described only recently (see \cite{Alexandrov:2010ca,Alexandrov:2014mfa,Alexandrov:2014rca}) and only on the type IIB side.
In this paper we will ignore them and restrict our attention to the simpler sector of D-instantons.

A D-instanton is characterized by a charge vector $\gamma = (p^\Lambda, q_\Lambda)$. On the type IIA side, it is integer valued
and labels the homology class $q_\Lambda \cA^\Lambda-p^\Lambda \cB_\Lambda \in H_3(\CY,\IZ)$ which contains the special Lagrangian submanifold
wrapped by a D2-brane.\footnote{The other branes existing in the type IIA formulations, D0 and D4, do not generate instanton contributions
because there are no 1- and 5-dimensional cycles on any Calabi-Yau threefold.} On the type IIB side, it labels instead an element of the
derived category of coherent sheaves \cite{Sharpe:1999qz,Douglas:2000gi}. Given the charge, one further introduces two other important objects:
\begin{itemize}
\item
the central charge function
\be
\label{defZ}
Z_\gamma(z) = q_\Lambda z^\Lambda- p^\Lambda F_\Lambda(z),
\ee
which appears as the central element in the supersymmetry subalgebra unbroken by the instanton;
\item
the generalized Donaldson-Thomas (DT) invariant (or simply the BPS index) $\hng$, which is an integer\footnote{In fact, the DT invariants
are piecewise constant functions on the moduli space parametrized by $z^a$. They jump across codimension one walls in this space, known
as lines of marginal stability, according to the wall-crossing formula of \cite{ks}. For the purpose of this work this phenomenon is irrelevant
and we can safely ignore this dependence on the moduli.}
appearing as a part of the topological data characterizing the Calabi-Yau $\CY$ and, in a sense, counts the instantons of given charge.
\end{itemize}
Then the leading contribution of the D-instanton of charge $\gamma$ to the metric has the following form \cite{Becker:1995kb}
\be
\label{d2quali}
\delta \de s^2\vert_{\text{D-inst}} \sim \,\hng\,
 e^{ -2\pi|Z_\gamma|/g_s
- 2\pi\I (q_\Lambda \zeta^\Lambda-p^\Lambda\tzeta_\Lambda)} .
\ee

In fact, one can do much better and incorporate D-instantons {\it exactly}, to all orders in $g_s$ and in the instanton expansion.
This is achieved using the twistorial description of QK manifolds \cite{Alexandrov:2008nk}.
The main idea behind this approach is that complicated constraints of the QK geometry are resolved in terms of
some holomorphic data on the twistor space $\cZ$. The latter is a canonical $\CP$-bundle over the QK manifold $\cM$, where the fiber describes
the triplet of almost complex structures $J^i$ satisfying the algebra of quaternions and thereby realizing the quaternionic structure of $\cM$.
The main advantage of the twistor space is that, in contrast to $\cM$, it is a \kahler manifold which carries in addition a {\it holomorphic contact
structure}. It is defined as the kernel of the canonical (1,0)-form $Dt$ on $\cZ$, where $t$ is a complex coordinate parametrizing the fiber.
This (1,0)-form is in turn determined by the $SU(2)$ part $\vec p$ of the Levi-Civita connection on $\cM$ as follows
\be
D t = \de t + p^+ - \I p^3 t + p^- t^2,
\label{canform}
\ee
where we used the chiral components of the connection, $p^\pm=-\hf\(p^1\mp \I p^2\)$.
Rescaling $Dt$, one can make from it a {\it holomorphic} one-form\footnote{In general, the rescaling factor may depend
holomorphically on the fiber coordinate $t$ and is different in different patches of an open covering of the twistor space,
which implies that the contact one form is not globally defined and has different local realizations $\cX^{[i]}$.
However, we will not need such generic construction which becomes relevant only after inclusion of NS5-brane instantons.}
\be
\cX = \frac{4}{\I t} \,e^{\phi}\, D t
\label{relcontform}
\ee
such that $\cX\wedge \(\de \cX\)^n$ is the non-vanishing holomorphic top form.
The rescaling function $\phi$ is called the {\it contact potential}. The properties of $\cX$ imply that locally,
by a proper choice of coordinates, it can always be trivialized as
\be
\cX = \de \ai{i} + \xii{i}^\Lambda \de \txii{i}_\Lambda,
\label{contform}
\ee
where the index $\scriptstyle{[i]}$ labels open patches of an atlas, $\cZ=\cup\, \cU_i$, and
$(\xii{i}^\Lambda,\txii{i}_\Lambda,\ai{i})$ is the set of {\it Darboux coordinates} in $\cU_i$.
These coordinates turn out to be the main object of interest in this construction because
knowing them as functions on the base $\cM$ and of the fiber coordinate $t$ is, in principle, equivalent to knowing the metric.
Indeed, combining \eqref{relcontform} and \eqref{contform}, one can find the contact potential $\phi$ and the $SU(2)$ connection $\vec p$,
which can then be used to compute the triplet of quaternionic two-forms $\vec \omega$.
They are defined by the almost complex structures, $\vec \omega(X,Y)=g(\vec J X,Y)$, and
the QK geometry requires that they are proportional to the curvature of the $SU(2)$ connection\footnote{The proportionality coefficient is
related to the (inverse) cosmological constant and affects only the overall scale of the metric.
We fix it by consistency with the perturbative metric \eqref{1lmetric}.}
\be
\vec \omega=-2\({\rm d} \vec p + \frac{1}{2}\,\vec p \times\vec p\).
\label{su2curvature}
\ee
On the other hand, the Darboux coordinates can also be used to get the almost complex structure $J^3$
so that, combining it with $\omega^3$, one arrives at the metric on $\cM$.
The details of this procedure are explained in appendix \ref{apap-metricgen}, and in the next section we apply it to extract
the D-instanton corrected metric on $\cM_H$.

Thus, to incorporate D-instantons in the twistor approach,
we should specify the Darboux coordinates on the twistor space of the HM moduli space $\cM_H$
taking into account their contributions. This was done in \cite{Alexandrov:2008gh,Alexandrov:2009zh}
and the resulting Darboux coordinates are determined in terms of functions $\Xigi(\varpi)$ which
satisfy the following system of integral equations
\be
\label{eqXigi}
\Xigi(\varpi)=\Thkl + \cR\(\varpi^{-1}\Zg-\varpi\bZg\)
+\frac{1}{8\pi^2}\sum_{\gamma'} \Om{\gamma'}\<\gamma,\gamma'\> \int_{\ellg{\gamma'}}\frac{\d \varpi'}{\varpi'}\,
\frac{\varpi+\varpi'}{\varpi-\varpi'}\,
\log\(1-\qrp e^{-2\pi \I \Xigipr(\varpi')}\).
\ee
Here $\Thkl= q_\Lambda \zeta^\Lambda - p^\Lambda\tzeta_\Lambda$, $\cR$ plays the role of a coordinate on the moduli space
(we will trade it later for the dilaton),
$\langle\gamma,\gamma'\rangle=q_\Lambda p'^\Lambda-q'_\Lambda p^\Lambda$ is the skew-symmetric product of charges,
$\ell_\gamma$ is the so-called BPS ray on $\CP$ joining $t=0$ and $t=\infty$
along the direction determined by the phase of the central charge
\be
\ell_\gamma=\{ t\, :\ Z_\gamma(z)/t\in \I \IR^-\},
\ee
and $\qr$ is a sign function on the charge lattice satisfying $\qr\qrp=(-1)^{\langle \gamma,\gamma'\rangle}\qrg{\gamma+\gamma'}$
and known as quadratic refinement (we set it to 1 for pure electric charges $\gamma=(0,q_\Lambda)$).
Given the functions $\Xigi(\varpi)$, the Darboux coordinates in the patch $\cU_\gamma$, which lies to the left from
the BPS ray $\ell_\gamma$, read as
\be
\label{exline}
\begin{split}
\xii{\gamma}^\Lambda =&\, \zeta^\Lambda + \cR \left(
\varpi^{-1} z^\Lambda - \varpi \, \bz^\Lambda\right) +
\frac{1}{8\pi^2}\sum_{\gamma'} \Om{\gamma'} p'^\Lambda \Iggp(\varpi) ,
\\
\txii{\gamma}_\Lambda =&\,
\tzeta_\Lambda
+\cR \left( \varpi^{-1} F_\Lambda - \varpi \, \bF_\Lambda \right)
+\frac{1}{8\pi^2}\sum_{\gamma'} \Om{\gamma'} q'_{\Lambda}\Iggp(\varpi)  ,
\\
\ai{\gamma}=&\,4\I c \log \varpi
-\hf\,\sigma-\frac{ \cR}{2}\,\(t^{-1} W-t \bar W\)
+\frac{\cR}{16\pi^2}\sum_{\gamma'} \Om{\gamma'}\(\varpi^{-1}Z_{\gamma'}+\varpi\bar{Z}_{\gamma'} \)\Iggp(0)
\\
&\,
-\frac{\I}{16\pi^3} \sum_{\gamma'} \Om{\gamma'} \int_{\ellg{\gamma'}}\frac{\d \varpi'}{\varpi'}\,
\frac{\varpi+\varpi'}{\varpi-\varpi'} \,L_{\qrp}\(e^{-2\pi \I \Xigi(\varpi')}\)
-\hf\, \xii{\gamma}^\Lambda(\varpi)\, \txii{\gamma}_\Lambda(\varpi),
\end{split}
\ee
where
\be
\label{defWnonThkl}
W(z) \equiv  F_\Lambda(z) \zeta^\Lambda - z^\Lambda \tzeta_\Lambda
\ee
and we introduced two functions\footnote{The first function is a variant of the Roger dilogarithm which satisfies
the famous {\it pentagon identity} and plays an important role in integrability \cite{Zagier-dilog}.}
\be
\begin{split}
L_\epsilon(z) =&\,  \Li_2 (\epsilon z) + \frac12\, \log z \log (1-\epsilon z),
\\
\Igg(\varpi)=&\, \int_{\ellg{\gamma}}\frac{\d \varpi'}{\varpi'}\,
\frac{\varpi+\varpi'}{\varpi-\varpi'}\,
\log\(1-\qr e^{-2\pi \I \Xigi(\varpi')}\).
\end{split}
\label{newfun}
\ee
These equations capture the effect of all D-instantons in an {\it exact} way.
The price to pay for this non-perturbative description is that it is somewhat implicit --- to get
corrections to the metric tensor, one needs to follow the procedure outlined in appendix \ref{apap-metricgen}.
The main obstacle on this way is the complicated nature of the integral equations \eqref{eqXigi}, which can be solved,
for generic set of charges, only perturbatively generating an instanton expansion.
This is the reason why below we restrict to a subset of charges which allows to avoid this problem.

A quantity, which is needed for evaluation of the metric and plays an important role in this story
\cite{Alexandrov:2011ac,Alexandrov:2014wca}, is the contact potential appearing in the relation \eqref{relcontform}.
It was explicitly evaluated in \cite{Alexandrov:2009zh}, again in terms of the solution of \eqref{eqXigi}, and is given by
\be
e^{\phi} = \frac{\cR^2}{4}\, K(z,\bz)-c
-\frac{ \I \cR}{32\pi^2}\sum\limits_{\gamma} \hng
\int_{\ellg{\gamma}}\frac{\d \varpi}{\varpi}\,
\( \varpi^{-1}\Zg -\varpi\bZg\)
\log\(1-\qr e^{-2\pi \I \Xigi(\varpi)}\).
\label{phiinstmany}
\ee
Its importance is partially explained by the fact that it can be identified with the dilaton field.
Then the formula \eqref{phiinstmany} can be considered as an equation which allows to find the coordinate $\cR$ as a function of the dilaton
and thereby to express all Darboux coordinates in terms of the standard fields of the type IIA formulation of string theory.

Finally, note that the Darboux coordinates \eqref{exline} carry a representation of the symplectic group.
Namely, $(\xii{\gamma}^\Lambda, \txii{\gamma}_\Lambda)$
transform as a vector under symplectic transformations, whereas the combination
$\ai{\gamma} +\hf\,\xii{\gamma}^\Lambda \txii{\gamma}_\Lambda$ is invariant. Besides, the contact potential \eqref{phiinstmany}
is also invariant. These properties ensure that the D-instanton corrections are consistent with symplectic invariance of
type IIA theory.

\subsection{Type IIB and mirror symmetry}
\label{subsec-IIB}

So far we dealt mostly with the HM moduli space in the type IIA formulation.
Its type IIB description can be obtained by applying {\it mirror symmetry} which requires that type IIA and type IIB
string theories compactified on mirror Calabi-Yau threefolds, and their moduli spaces in particular, are the same.
However, $\cM_H$ in type IIB comes with its own set of natural coordinates. They are different from those used above
and adapted to the action of the S-duality group $SL(2,\IZ)$, which is a manifest symmetry of the type IIB formulation.
Thus, to apply mirror symmetry in practice we need to know the relation between the type IIB fields
and the ones described in section \ref{subsec-pertHM}. Such relation is known as {\it mirror map}. At classical level
it has been found in \cite{Bohm:1999uk} and quantum corrections, including various instanton effects, have been included in
\cite{Alexandrov:2009qq,Alexandrov:2012bu,Alexandrov:2013mha}.

In this paper we will not need these general results.
For our purposes it will be sufficient to restrict to the mirror map for the field $\cR$, complex structure moduli $z^a$ and
RR-fields $\zeta^\Lambda$ in the presence of D-instantons with vanishing magnetic charge $p^\Lambda$.
In this approximation the mirror map for these fields turns out to coincide with the classical one and is given by
\be
\cR=\frac{\tau_2}{2}\,
\qquad
z^a=b^a+\I t^a,
\qquad
\zeta^0 =\tau_1 ,
\qquad
\zeta^a=-(c^a-\tau_1 b^a).
\label{mirmap}
\ee
The type IIB fields appearing here on the r.h.s. transform in the following way under an S-duality transformation
$\(\begin{array}{cc} a & b \\ c & d \end{array}\)\in SL(2,\IZ)$:
\be
\tau\to\frac{a\tau+b}{c\tau+d},
\qquad
t^a\to |c\tau+d| \, t^a,
\qquad
\(\begin{array}{c} c^a \\ b^a \end{array}\)\to \(\begin{array}{cc} a & b \\ c & d \end{array}\)\(\begin{array}{c} c^a \\ b^a \end{array}\),
\label{Sdualtr}
\ee
where we combined the inverse 10-dimensional string coupling $\tau_2=1/g_s$ with the RR-field $\tau_1$ into
an axio-dilaton $\tau=\tau_1 + \I \tau_2$.
We will use these relations in section \ref{sec-sing} to extract the strong coupling behavior of certain contributions to the HM metric.

%-------------------------------------------------------------------------------------
\section{D-instanton corrected metric}
\label{sec-metric}

In this section we will provide an explicit expression for the D-instanton corrected HM metric, deriving it from the twistorial construction
presented in section \ref{subsec-Dinst}. We will relegate most of intermediate equations and technical details
to appendix \ref{ap-metric}, trying to concentrate here on conceptual issues.

In fact, we are not able to compute {\it exactly}
the metric which includes {\it all} D-instanton corrections because it is not possible to solve explicitly the integral equations \eqref{eqXigi},
which are at the heart of this construction. Therefore, we impose an additional condition that
all charges are mutually local, i.e.
\be
\langle \gamma,\gamma'\rangle=0.
\label{mutloc}
\ee
This condition can be interpreted in two ways.
On one hand, it is satisfied if we include only electrically charged D2-instantons which have charges
with vanishing magnetic component $\gamma=(0,q_\Lambda)$. Any other set of charges solving \eqref{mutloc}
can be rotated to this one by a symplectic transformation. Nevertheless, it is useful to work in generic frame because
it allows to check the symplectic invariance of the final result, which is done in appendix \ref{ap-sympl}.
On the other hand, the condition \eqref{mutloc} can be viewed as a reduction to the one-instanton approximation
because it effectively kills all multi-instanton terms in the expressions for Darboux coordinates \eqref{exline}.
This provides another justification for not setting magnetic charges to zero at once.

The assumption \eqref{mutloc} liberates us from the necessity to solve any equations since it reduces \eqref{eqXigi}
to an explicit and simple expression
\be
\label{eqXigi-ml}
\Xigi(\varpi)=\Thkl + \cR\(\varpi^{-1}\Zg-\varpi\bZg\).
\ee
This is the crucial simplification. From this point no more approximations or assumptions need to be made
to compute the metric explicitly.
The general procedure to extract it from Darboux coordinates on the twistor space is presented in appendix \ref{apap-metricgen}.
The idea is just to apply this procedure to the system described by eqs. \eqref{exline} and \eqref{phiinstmany}.

However, first, we should translate the Darboux coordinates to the patch around $t=0$, the north pole of $\CP$, which we denote by $\cU_+$.
This can be done by performing a holomorphic contact transformation, i.e. a change of Darboux coordinates preserving the contact one-form \eqref{contform},
which removes most of singularities at $t=0$ and leaves only those which are admitted by the condition \eqref{expDc}: $\xi^\Lambda$ can have
a simple pole, $\txi_\Lambda$ should be regular, and $\alpha$ has only a logarithmic singularity controlled by the one-loop correction $c$.
Such contact transformation is given by
\be
\begin{split}
\xii{+}^\Lambda &=  \xii{\gamma}^\Lambda +\p_{\txi_\Lambda }\Hij{+\gamma},
\\
\txii{+}_\Lambda &=  \txii{\gamma}_\Lambda - \p_{\xi^\Lambda } \Hij{+\gamma},
\\
\ai{+} &=  \ai{\gamma}-\Hij{+\gamma}+ \xii{+}^\Lambda \p_{\xi^\Lambda}\Hij{+\gamma},
\end{split}
\label{QKgluing}
\ee
where the holomorphic function $\Hij{+\gamma}$ was found in \cite{Alexandrov:2009zh} to have the following form
\be
\label{gensymp}
\Hij{+\gamma}=  F(\xii{+}) +\Gg(\xii{+},\txii{\gamma}).
\ee
Here the second term is a complicated, but irrelevant function for us because, as was shown in \cite{Alexandrov:2009zh},
it affects only $O(t^2)$ terms in the Laurent expansion of the Darboux coordinates.
Thus, we can safely ignore it for our purposes, and this allows to replace $\xii{+}^\Lambda$ on the r.h.s. of \eqref{QKgluing} by
$\xii{\gamma}^\Lambda$.

After this, it is straightforward to compute first few coefficients in the Laurent expansion around $t=0$
which can be found in \eqref{LaurentDarboux}. Substituting them into \eqref{connection}, one finds the components of the SU(2) connection $\vec p$,
see \eqref{SU(2)connect}. This connection in turn can be used to get the quaternionic 2-form $\omega^3$ via \eqref{Kform}. The result
is given in \eqref{om3}. In all these results we extensively used notations defined in appendix \ref{notations}
and, as for the perturbative metric, denoted by $r$ the exponential of the dilaton identified with the contact potential.

The next step is to write down explicitly the basis of (1,0) forms in the almost complex structure $J^3$.
It is given by \eqref{defPi}, but can be further simplified. First, since $\pi^a=\de z^a$, one can drop all terms proportional to this one-form
in other basis elements. Furthermore, it turns out to be convenient to add to $\tilde\pi_\alpha$ the term
$-\frac{\I}{2}\,\xii{+}^{\Lambda,0} \tilde\pi_\Lambda$.
As a result, one arrives at the following basis
\be
\begin{split}
\de z^a, &
\\
\cY_\Lambda = &\,
\de\tzeta_\Lambda-F_{\Lambda\Sigma}\de\zeta^\Sigma
-\frac{1}{8\pi^2}\sum\limits_{\gamma} \hng{} \(q_{\Lambda}-p^\Sigma F_{\Lambda\Sigma}\)\de \Igg{},
\\
\Sigma=&\, \de r +2c\,\de \log\cR-\frac{\I}{16\pi^2}\sum\limits_{\gamma} \hng{} \(\cR\Zg{} \de\Igp-\Igm \de\(\cR\bZg{}\)\)
\\
&\,
+\frac{\I}{4}\(\de \sigma+\tzeta_\Lambda\de \zeta^\Lambda-\zeta^\Lambda\de \tzeta_\Lambda\).
\label{holforms}
\end{split}
\ee

The final step is the most cumbersome. It requires to rewrite the quaternionic 2-form $\omega^3$ \eqref{om3}
in the basis of 1-forms \eqref{holforms} and their complex conjugates, so that it takes the form similar to \eqref{metom}.
This is a straightforward, although lengthy procedure which is the subject of appendix \ref{ap-om3}.
The final result can be found in \eqref{final-om3} and immediately leads to
the following expression for the D-instanton corrected metric:
\bea
\de s^2&=&
\frac{2}{r^2} \(1-\frac{2r}{\cR^2\Uin}\)(\de r)^2
+\frac{1}{32r^2\(1-\frac{2r}{\cR^2\Uin}\)}\(\de \sigma +\tzeta_\Lambda \de \zeta^\Lambda-\zeta^\Lambda\de \tzeta_\Lambda+\cV \)^2
\nn\\
&&
+\frac{\cR^2}{2r^2}\,|z^\Lambda\cY_\Lambda|^2
+\frac{1}{r\Uin}\left|  \cY_\Lambda \Min^{\Lambda\Sigma}\bvl_\Sigma-\frac{\I\cR}{2\pi}\, \sum_\gamma \Om{\gamma}\cW_\gamma\de Z_\gamma\right|^2
\nn\\
&&
-\frac{1}{r}\,\Min^{\Lambda\Sigma}\(\cY_\Lambda
+\frac{\I\cR}{2\pi}\sum_\gamma \Om{\gamma} V_{\gamma\Lambda}\rIgamp{\gamma}\(\de Z_{\gamma}-\Uin^{-1}Z_\gamma\p K\)\)
\nn\\
&&\qquad
\times\(\bar\cY_\Sigma-\frac{\I\cR}{2\pi}\sum_{\gamma'} \Om{\gamma'} \bV_{\gamma'\Sigma}
\rIgamm{\gamma'}\(\de \bZ_{\gamma'}-\Uin^{-1}\bZ_{\gamma'}\bar\p K\)\)
\nn\\
&&
+\frac{\cR^2 K}{r}\Biggl( \cK_{a\bar b}\de z^a \de\bz^b
-\frac{1}{(2\pi K\Uin)^2}\left|\sum_\gamma \hng{}Z_\gamma \cW_\gamma\right|^2 |\p K|^2
\Biggr.
\nn\\
&&\Biggl.\qquad
+\frac{1}{2\pi K}\sum_\gamma \hng{} \rIg\left|\de Z_\gamma-\Uin^{-1}Z_\gamma\p K\right|^2
\Biggr).
\label{mett1}
\eea

This is the main result of this work.
Several comments about it are in order.
\begin{itemize}
\item
To keep the expression for the metric as simple as possible, we used several notations introduced in appendix \ref{notations}:
\begin{itemize}
\item
$\Ig{}$ with various upper indices denote the twistorial integrals \eqref{newfun-expand}, which all can be
evaluated in terms of series of Bessel functions;
\item
$V_{\gamma\Lambda}$ and $\vv_\Lambda$ are the vectors \eqref{defVgam} and \eqref{notvec};
\item
$\Min^{\Lambda\Sigma}$ is the inverse of the matrix \eqref{defmatM};
\item
$\Uin$ is the function \eqref{defUin},
which can be thought of as an instanton corrected version of the \kahler potential;
\item
$\cW_\gamma$ is a function on the charge lattice defined in \eqref{cWdef};
\item
and finally $\cV$ is the one-form \eqref{conn}
generalizing the \kahler connection \eqref{kalcon} appearing in the perturbative metric.
\end{itemize}

\item
As was promised, the expression \eqref{mett1}, although somewhat non-trivial, is rather explicit. However, there are two
implicit ingredients which still may require to make an instanton expansion. First of all, this is the inverse matrix $\Min^{\Lambda\Sigma}$.
Only in some particular cases it can be found without involving any expansion. Secondly, this is the coordinate $\cR$ which
should be viewed as a function of other coordinates on the moduli space. This function
is defined only implicitly by the expression for the dilaton \eqref{phiinstmany},
which in our notations takes the following form
\be
r = \frac{\cR^2}{4}\, K-c
-\frac{ \I\cR}{32\pi^2}\sum\limits_{\gamma}\hng{} \(Z_\gamma\Igp+\bZ_\gamma\Igm\) .
\label{dil}
\ee
\item
Since the HM metric is derived using the assumption on the D-brane charges which has a
symplectic invariant form, it is expected to be symplectic invariant itself. However, this symmetry is not explicit
in the form given in \eqref{mett1}. In fact, it is not explicit in the expression \eqref{1lmetric} for the perturbative metric either.
In that case it is actually not so difficult to bring the metric to a manifestly symplectic invariant form, see for instance \cite[Eq.(3.12)]{Alexandrov:2011va}.
In our case this is a harder task, mainly due to the presence of the matrix $\Min^{\Lambda\Sigma}$. Nevertheless, in appendix \ref{ap-sympl}
we address this issue and prove that the metric is indeed symplectic invariant.

\item
As was noticed in the beginning of this section, our result is valid in the two cases:
either we include only a subset of all possible D-instantons with charges satisfying \eqref{mutloc},
in which case the metric is exact, or one considers all charges but restricts to the one-instanton approximation.
This approximation is effectively equivalent to dropping all terms non-linear in DT invariants $\hng$.
In this case the metric can be further simplified. In particular, one can explicitly invert the matrix $\Min^{\Lambda\Sigma}$
and solve \eqref{dil} for $\cR$ as a function of other coordinates.

\item
The instanton corrections break the nice two-stage fibration structure \eqref{doublefib2} of the perturbative metric.
This happens, first of all, due to the fact that the metric on the subspace parametrized by the complex structure moduli
acquires a dependence on the RR-fields. Moreover, the dilaton is now not factorized anymore, but
non-trivially combined with both $z^a$ and $(\zeta^\Lambda,\tzeta_\Lambda)$ due to the appearance of terms proportional to $\de r$
in the holomorphic one-form $\cY_\Lambda$, see \eqref{expand-Y}.
The property, which however remains true, is that the HM moduli space is still a circle bundle with the fiber parametrized by the NS-axion,
\be
\label{quant-fib}
\begin{array}{rl}
S^{1}_\sigma\ \longrightarrow\ &\cM_H
 \\
&\  \downarrow
\\
& \cB_{r,z,\zeta,\tzeta}^{\rm inst}\, .
\end{array}
\ee
A non-trivial feature of the metric \eqref{mett1} is that the connection defining this bundle,
$\tzeta_\Lambda \de \zeta^\Lambda-\zeta^\Lambda\de \tzeta_\Lambda+\cV$,
does not have a component along the dilaton, see \eqref{conn}.
This might be related to the obstructions on the quantum corrections coming from the relation \eqref{curvS}.
\end{itemize}

\section{Universal hypermultiplet}
\label{sec-UHM}

\subsection{Tod ansatz}
\label{subsec-Tod}

A very important particular case of our story corresponds to compactification on a rigid Calabi-Yau manifold, i.e. the one
with $h^{2,1}(\CY)=0$ which therefore does not have complex structure moduli.
The vanishing of the Hodge number implies that the HM sector of type IIA string theory compactified on $\CY$ consists
only from one hypermultiplet, appearing in the literature under the name of the universal hypermultiplet \cite{Strominger:1997eb}.
Its moduli space is a four-dimensional QK manifold. In four dimensions the QK condition is more explicit than in higher dimensions
and implies that the manifold should be an Einstein space with a non-vanishing cosmological constant and a self-dual Weyl curvature.

Once the effects NS5-brane instantons are ignored, the HM moduli space is guaranteed to have at least one continuous isometry,
which acts by constant shifts of the NS-axion $\sigma$.
Self-dual Einstein spaces with such an isometry admit a rather explicit description: by a proper choice of coordinates,
their metric can always be written in the form of the Tod ansatz parametrized by one real function \cite{MR1423177}.
The ansatz reads
\be
\de s^2
=-\frac{3}{\Lambda}\[\frac{P}{\rho^2}\(\de \rho^2+4 e^T\de z\de \bz\)+\frac{1}{P\rho^2}\(\de\theta+\Theta\)^2\],
\label{met-Toda}
\ee
where $T$ is a function of $(\rho,z,\bz)$ and is independent of $\theta$ parametrizing the direction of the isometry.
Furthermore,
\bea
P &=& 1-\hf\, \rho\p_\rho T,
\\
\de \Theta &=& \I\(\p_z P\de z-\p_{\bz} P\de\bz\)\wedge \de \rho-2\I \p_\rho(P e^T)\de z\wedge \de \bz,
\label{deThet}
\eea
whereas the Einstein self-duality condition of the metric is encoded in the Toda differential equation
to be satisfied by the function $T$,
\be
\p_z\p_{\bz} T+\p_\rho^2 e^T=0.
\label{eq-Toda}
\ee

This description was at the origin of many attempts to compute the instanton corrected metric on $\cM_H$ because it implies
that all instanton corrections can be encoded just in one function, the Toda potential $T$. To extract them, it is sufficient
to find a proper solution of the Toda equation. Although in the one-instanton approximation this strategy was very successful
\cite{Ketov:2001gq,Ketov:2001ky,Ketov:2002vr,Saueressig:2005es,Theis:2014bia}\footnote{A similar strategy can be applied to derive
NS5-brane instantons as well \cite{Alexandrov:2006hx}, because generic 4d QK manifolds
can be parametrized by solutions of another, more complicated non-linear differential equation, which replaces the Toda equation
in the absence of the isometry \cite{Przanowski:1984qq}.},
the results obtained beyond this approximation
are often not reliable because of additional unjustified simplifications typically imposed on the ansatz for $T$ to fix ambiguities
of integration and to avoid complications of the full non-linear problem.

Given the twistorial construction of D-instantons, we do not need to solve any differential equations anymore.
In principle, this construction should provide us automatically with a solution of the Toda equation which incorporates all
D-instanton corrections. Furthermore, in \cite{Alexandrov:2009vj} a dictionary between the twistorial quantities and
those of the Tod ansatz, which should be sufficient to extract such a solution, was found. It is given by the following relations
\be
\rho=e^\phi,
\qquad
z=\frac{\I}{2}\, \txii{+}_0,
\qquad
\theta= -\frac{1}{8}\, \sigma,
\qquad
T=2\log (\cR/2).
\label{dict-Toda}
\ee
In the next subsection we will show that the HM metric computed in the previous section,
specialized to the four-dimensional case, reproduces the ansatz \eqref{met-Toda}, and the resulting Toda potential and coordinates
are consistent with the relations \eqref{dict-Toda}.

\subsection{The metric and Toda potential}
\label{subsec-Todapot}

To write the metric \eqref{mett1} for the universal hypermultiplet, we note that in this case the indices $\Lambda,\Sigma,\dots$
take only one value, whereas quantities with indices $a,b,\dots$ do not simply exist. Correspondingly, we drop
the remaining index 0 on $n$-dimensional vectors such as charges and RR-fields and denote them simply as
$\gamma=(p,q)$ and $(\zeta,\tzeta)$.
Furthermore, for rigid Calabi-Yau manifolds the holomorphic prepotential is a quadratic monomial \cite{Bao:2009fg}
\be
F(X) = \frac{\lambda}{2}\, X^2,
\qquad\quad
\lambda\equiv \lambda_1 -\I \lambda_2=\frac{\int_\cB\Omega}{\int_\cA\Omega}\, ,
\label{prepUHM}
\ee
where $\lambda$ is a fixed complex number, given by the ratio of periods of the holomorphic 3-form $\Omega\in H^{3,0}(\CY)$
over an integral symplectic basis $(\cA,\cB)$ of $H_3(\CY,\IZ)$, with $\lambda_2>0$.
As a result, for the quantities characterizing the ``special \kahler geometry"\footnote{We put quotes because in this case
they describe an empty space.}, one finds $N=K=2\lambda_2$
and $Z_\gamma= V_{\gamma}=q-\lambda p $.
The other important quantities, including the potential $\Uin$ and the one-forms $\cY$ and $\cV$, can be found in \eqref{Ab-UHM}.
Plugging them into \eqref{mett1}, the metric reduces to
\be
\de s^2=
\frac{2}{r^2}\[\(1-\frac{2r}{\cR^2\Uin}\) \((\de r)^2+\frac{\cR^2}{4}\,|\cY|^2\)
+\frac{1}{64}\(1-\frac{2r}{\cR^2\Uin}\)^{-1}\(\de \sigma +\tzeta \de \zeta-\zeta\de \tzeta+\cV \)^2\].
\label{mett-UHM}
\ee
Finally, it should be noted that the relation between $\cR$ and the dilaton $r=e^\phi$ is provided as usual by \eqref{dil}.

Comparing the resulting metric with the Tod ansatz \eqref{met-Toda}, one finds that they match perfectly provided one takes
$\Lambda=-3/2$, uses the identifications \eqref{dict-Toda}\footnote{We remind that $\cY=\de\txii{+}_{0}$, which implies
the identification $\de z=\frac{\I}{2}\,\cY$. }, and in addition ensures that
\begin{itemize}
\item
$
\p_\rho T=4\(\cR^2\Uin\)^{-1};
$
\item
the connection $\cV$ satisfies the analogue of \eqref{deThet} (see \eqref{secondcond});
\item
and the potential $T=2\log (\cR/2)$ fulfils the Toda equation.
\end{itemize}
In appendix \ref{ap-Toda} we prove that all these conditions do hold, and thus the metric we computed satisfies the constraints
of four-dimensional \qk geometry. This might be considered as a non-trivial test on the general metric \eqref{mett1}.

One of byproducts of our analysis is that we found an {\it exact} solution of the Toda equation.
Unfortunately, it is given only implicitly: it turns out to be encoded in the two {\it non-differential} equations.
One of them is the formula for the dilaton \eqref{dil}
and the second equation is \eqref{Lexptxi}, where $\txii{+}_0$ should be replaced by $-2\I z$.
The former allows to find the Toda potential $T=2\log (\cR/2)$
as a function of $r\!=\!\rho$ and the RR-fields $(\zeta,\tzeta)$, whereas the latter relates these fields to the complex coordinate $z$.
Choosing the electric frame for the set of mutually local charges and evaluating the integrals explicitly in terms of Bessel functions,
these two equations can be written as follows
\be
\begin{split}
e^T= &\,\frac{1}{2\lambda_2}(\rho +c)
-\frac{ e^{T/2}}{4\pi^2\lambda_2}\sum\limits_{q>0} \bOm{(0,q)}q\,
\cos\bigl(2\pi  q\zeta\bigr)
 K_1 \bigl(8\pi q e^{T/2}\bigr),
\\
z =&\, \frac{\I}{2}\,(\tzeta-\lambda \zeta)
+\frac{1}{4\pi^2}\sum\limits_{q>0} \bOm{(0,q)} q\,
 \sin\bigl(2\pi q\zeta\bigr) K_0  \bigl(8\pi q e^{T/2}\bigr) ,
\end{split}
\label{dilUHM}
\ee
where we used the so-called rational DT invariants
$
\bOm{\gamma} = \sum_{d\vert\gamma} \frac{1}{d^2} \, \Om{\gamma/d}
$, which appear here as free parameters of the solution.
It is clear that if one solves these equation by expanding in powers of instantons, the Toda potential will
be represented as a power series in Bessel functions, which is similar to the solution found in \cite{Theis:2014bia}.

In fact, the twistorial formalism allows to get the Toda potential which encodes {\it all} D-instanton corrections, and not only
the electrically charged ones. This is because the identifications \eqref{dict-Toda} hold in this more general case as well.
Such Toda potential will be again encoded in two non-differential equations similar to \eqref{dilUHM} and
corresponding to the formulas for the dilaton \eqref{phiinstmany} and
for the Fourier coefficient of the Darboux coordinate $\txi$ \eqref{exline}, where
no restriction on charges is imposed anymore.
However, the difference will be that the integrals appearing in these formulas cannot be evaluated explicitly,
because now the function $\Xi_\gamma(t)$ is not the simple polynomial \eqref{eqXigi-ml}, but satisfies the system of integral equations \eqref{eqXigi}.
Nevertheless, it is easy to solve this system perturbatively and generate an instanton expansion for $\Xi_\gamma(t)$ and
subsequently for the Toda potential. We do not give such an expansion here, but just note that
already at second order the solution is given by a complicated double integral,
which cannot be reduced to a product of Bessel functions.

%-------------------------------------------------------------------------------------
\section{Curvature singularity in the presence of D-instantons}
\label{sec-sing}

\subsection{Equation for singularity}

As was mentioned in the end of section \ref{subsec-pertHM}, after inclusion of the one-loop correction, if $\chi_\CY>0$,
the HM metric acquires a curvature singularity at $r=-2c$. The natural question is what happens with this singularity
once one adds D-instanton contributions. In this section we will try to answer this question, at least in the case of
electrically charged D-instantons where the metric \eqref{mett1} holds to all orders.

First of all, comparing the metrics \eqref{mett1} and \eqref{1lmetric}, one observes that the factor $r+2c$,
appearing in front of the kinetic terms for the dilaton and the NS-axion, is promoted now to
\be
\hf\,\cR^2 \Uin-r.
\label{singfactor}
\ee
Therefore, it is natural to expect that, equating this expression to zero, one obtains an equation
which determines a hypersurface in $\cM_H$ representing the singularity of the D-instanton corrected metric.
Substituting explicit expressions for $\Uin$ \eqref{defUin} and $r$ \eqref{dil}, one arrives at the following condition
\bea
&&
K+\frac{4c}{\cR^2}
+\frac{1}{\pi}\sum_\gamma \hng{}\(\frac{\I}{8\pi}\(\Zg{}\Igp+\bZg{}\Igm\)-|\Zg{}|^2\rIg\)
\label{singeqB}\\
&& +\frac{1}{8\pi^2} \(\sum_{\gamma}\Om{\gamma} \( \Zg{}\rIgp+\bZg{}\rIgm\) V_{\gamma\Lambda}\)
\Min^{\Lambda\Sigma}\sum_{\gamma'}\(\Om{\gamma'} \( Z_{\gamma'}\rIgamp{\gamma'}+\bZ_{\gamma'}\rIgamm{\gamma'}\)  \bV_{\gamma'\Sigma}\)
=0.
\nn
\eea

There is actually another way to get this equation, which also reveals what becomes singular from the geometric point of view
when one approaches the singularity.
It was noticed in \cite{Alexandrov:2009qq}, that the curvature singularity appearing at one-loop corresponds
to the degeneracy of the basis of (1,0)-forms. Let us see when this can happen. From the explicit form of this basis \eqref{holforms},
it follows that $\de z^a$, $\cY_\Lambda$ and $\Im\Sigma$ are always linearly independent. Thus, it can degenerate only if
\bea
\Re\Sigma = 0 \qquad \mod \de z^a,\ \cY_\Lambda.
\label{degcond}
\eea
For instance, at perturbative level this condition becomes
\bea
\de e^\phi + 2 c \de \log \cR = 0 \qquad \mod \de z^a,\ \cY_\Lambda,
\label{1lholforms}
\eea
and upon using the one-loop relation between the coordinate $\cR$ and the dilaton
\bea
\cR = 2K^{-1/2} \sqrt{r+c},
\eea
one indeed gets back the standard perturbative result $r=-2c$.
Similarly, one can show that, applied to the instanton corrected basis of (1,0)-forms \eqref{holforms},
the condition \eqref{degcond} generates the equation \eqref{singeqB}.\footnote{In fact, this procedure was already
attempted in \cite{Alexandrov:2009qq}, but due to a simple computational mistake, the result presented there is wrong.}

Now the crucial question to be understood is whether the equation \eqref{singeqB} has any solutions.
In this respect it is important to point out that, in contrast to the perturbative result,
it depends not only on the dilaton, but on other coordinates as well, and is written more naturally in terms of $\cR$.
The latter is in a sense a more fundamental quantity because it is related to the 10-dimensional string coupling, see \eqref{mirmap},
whereas $r=e^\phi$ is a derived quantity and it is possible that its range of values allowed by \eqref{dil} is less than the positive half-axis.

Since for small string coupling, where $\cR\to\infty$, all terms in \eqref{singeqB} are suppressed comparing to the first one, which is always positive,
the equation will have a solution if there is a region in the moduli space where its l.h.s. is negative.
This might happen only at finite string coupling. Therefore, we need to understand the behavior of \eqref{singeqB} in such deep quantum regime.
This is possible when we have S-duality at our disposal which relates the weak and strong coupling regions.
Fortunately, this is the case in our situation because the sector of quantum corrections
we considered is S-duality invariant. In fact, this feature was used to find these corrections in the original work \cite{RoblesLlana:2006is}.
However, to make this symmetry explicit and to exploit it, one needs to pass to the mirror type IIB formulation.
In the next subsection we show how this can be done and analyze the behavior of \eqref{singeqB} under S-duality transformations.

\subsection{Equation for singularity and S-duality}

In the type IIB formulation electrically charged D-instantons correspond to contributions from D(-1) and D1-branes.
The former are point-like objects having only one non-vanishing charge $q_0$,
and the latter wrap two-dimensional cycles on the Calabi-Yau labeled by $q_a$.
Besides these exponential corrections in $g_s$, the metric receives
$\alpha'$-corrections through the holomorphic prepotential $F(X)$:
there is a perturbative correction and exponential contributions coming from worldsheet instantons.
The resulting HM metric should carry an isometric action of the S-duality group $SL(2,\IZ)$, which mixes
the perturbative $\alpha'$-correction with D(-1)-instantons and worldsheet with D1-instantons.
However, this symmetry is not manifest neither in the expression for the metric \eqref{mett1},
nor in the twistorial construction of section \ref{subsec-Dinst}, as they are adapted to the type IIA formulation.
The twistorial construction of D1-D(-1)-instantons has in fact been put in a manifestly S-duality invariant form
in \cite{Alexandrov:2009qq,Alexandrov:2012bu}, which can be seen as an indirect proof of the invariance of the metric.
This also implies that the equation for singularity should be S-duality invariant as well.

To see this explicitly, one needs to rewrite the equation \eqref{singeqB} in the type IIB variables using the mirror map \eqref{mirmap}
and perform a Poisson resummation. Below we will do this for the sector which includes D(-1)-instantons only.
Thus, we neglect contributions from worldsheet and D1-instantons. This is justified not only by simplifications which happen in this approximation,
but also by the fact that the one-loop $g_s$-correction giving rise to the singularity is a part of the same $SL(2,\IZ)$ multiplet as 
D(-1)-instantons. Therefore, one could hope that already contributions from D(-1)-branes are sufficient to
resolve the singularity.

Thus, in the following we consider D-brane charges with only one non-vanishing component $q_0$.
In this case the DT invariant is independent of $q_0$ and coincides with the Euler characteristic of the Calabi-Yau threefold,
$\Om{q_0}=\chi_\CY=-\chi_{\CYm}$ where $\CYm$ is the Calabi-Yau on which type IIB string theory is compactified and which is mirror
to $\CY$ used in type IIA compactification.
Since we drop the contribution of worldsheet instantons, the holomorphic prepotential reads \cite{Candelas:1990rm}
\be
F(X)= -\kappa_{abc}\,\frac{X^a X^b X^c }{6X^0} +\frac{\I\zeta(3)\chi_\CYm}{16\pi^3}\, (X^0)^2,
\label{prep}
\ee
where the first term is the classical contribution determined by the intersection numbers $\kappa_{abc}$ of 4-cycles,
whereas the second term is a perturbative $\alpha'$-correction.
This prepotential leads to the exponential of the \kahler potential given by
\be
K = 8V-\frac{\zeta(3)\chi_\CYm}{4\pi^3},
\ee
where $V=\frac{1}{6}\, \kappa_{abc} t^a t^b t^c$ is the Calabi-Yau volume and we used the relation $z^a=b^a+\I t^a$
from the mirror map \eqref{mirmap}.
Substituting these data in the singularity equation \eqref{singeqB}, it takes the following form
\be
8V-\frac{\chi_\CYm}{2\pi}\, \BI+\frac{\chi_\CYm^2}{8\pi^2}\,\Min^{00}\CI^2=0,
\label{singeq-D1}
\ee
where we introduced
\be
\begin{split}
\BI =&\,\sum_{q_0\ne 0} q_0\(\frac{\I}{4\pi}\(\Igp+\Igm\)-2q_0\rIg\) +\frac{\zeta(3)}{2\pi^2}-\frac{1}{24\cR^2},
\\
\CI =&\, \sum_{q_0\ne 0} q_0^2\( \rIgp+\rIgm\).
\end{split}
\label{defS12}
\ee

Furthermore, in this approximation it is possible to find an explicit expression for $\Min^{00}$.
Indeed, for the holomorphic prepotential \eqref{prep}, the matrix \eqref{defmatM} reads
\be
\Min_{\Lambda\Sigma} = \Ncl_{\Lambda\Sigma}-\frac{\AI \chi_\CYm}{2\pi}\,\,\delta_\Lambda^0\delta_\Sigma^0,
\label{MinD-1}
\ee
where
\be
\AI=\frac{\zeta(3)}{2\pi^2}-\sum_{q_0\ne 0} q_0^2\rIg
\label{defS3}
\ee
and
\be
\Ncl_{\Lambda\Sigma}=\(\begin{array}{cc}
-4V+2\kappa_{abc}b^a b^b t^c  & \ -2\kappa_{abc} b^b t^c
\\
-2\kappa_{abc} b^b t^c & \ 2\kappa_{abc} t^c
\end{array}\)
\ee
is the same as the matrix $N_{\Lambda\Sigma}$ but defined only by the first classical term in the prepotential \eqref{prep}.
Its inverse can be computed in terms of the matrix $G_{ab}=\frac{1}{2V}\, \kappa_{abc} t^c$ which is supposed to be invertible.
Then one finds
\be
(\Ncl)^{\Lambda\Sigma}=-\frac{1}{4V} \(\begin{array}{cc}
1 &\  b^a
\\
b^a  &\  b^a b^b - (G^{-1})^{ab}
\end{array}\).
\ee
On the other hand, it is easy to show that
\be
\Min^{\Lambda\Sigma}=(\Ncl)^{\Lambda\Sigma}+
\frac{\AI \chi_\CYm}{2\pi}\,\frac{(\Ncl)^{\Lambda 0}(\Ncl)^{\Sigma 0}}{1-\frac{\AI \chi_\CYm}{2\pi}\,(\Ncl)^{00}}\,.
\ee
In particular, this implies
\be
\Min^{00}=\frac{(\Ncl)^{00}}{1-\frac{\AI \chi_\CYm}{2\pi}\,(\Ncl)^{00} }=-\frac{1}{4V+\frac{\AI \chi_\CYm}{2\pi} }\, .
\label{resM00}
\ee
Substituting this result into \eqref{singeq-D1}, the singularity equation can be brought to the following form
\be
1+\frac{\chi_\CYm}{16\pi V}\(2\AI-\BI\)
-\frac{\chi_\CYm^2}{(16\pi V)^2}\(2\AI\BI+\CI^2\)=0.
\label{singeq-red}
\ee

The last step is to perform the Poisson resummation of the three functions $\BI$, $\CI$ and $\AI$.
This is done is appendix \ref{ap-resum} and gives
\bea
\BI&=&\,   \frac{1}{4\pi^2}\mathop{{\sum}'}\limits_{m,n \in \IZ}\(\frac{1}{|m\tau+n|^3}-\frac{3(m\tau_2)^2}{|m\tau+n|^5}\),
\nn
\\
\CI &=&\,  - \frac{3}{4\pi^2}\mathop{{\sum}'}\limits_{m,n \in \IZ}\frac{m\tau_2(m\tau_1+n)}{|m\tau+n|^5}\, ,
\label{AIBICI}
\\
\AI&=&\,  \frac{1}{4\pi^2}\mathop{{\sum}'}\limits_{m,n \in \IZ}\(\frac{1}{|m\tau+n|^3}-\frac{3(m\tau_2)^2}{2|m\tau+n|^5}\),
\nn
\eea
where prime means that the sum goes over all pairs of integers except $(m,n)=(0,0)$.
Introducing the non-holomorphic Eisenstein series
\be
E_{3/2}(\tau)=\mathop{{\sum}'}\limits_{m,n \in \IZ}\frac{\tau_2^{3/2}}{|m\tau+n|^3},
\ee
which is a modular invariant function, and plugging these results into \eqref{singeq-red},
one finally arrives at the equation for singularity in the type IIB variables, which makes
its modular properties manifest,
\be
1+\frac{\chi_\CYm\, E_{3/2}(\tau)}{64\pi^3 V\tau_2^{3/2}}
-\frac{2\chi_\CYm^2}{\bigl(64\pi^3 V\tau_2^{3/2}\bigr)^2}\[E_{3/2}^2(\tau)
-\frac{9}{4} \mathop{{\sum}'}\limits_{m,n \in \IZ}\mathop{{\sum}'}\limits_{m',n' \in \IZ}\frac{\tau_2^5(mn'-nm')^2}{|m\tau+n|^5|m'\tau+n'|^5}\]=0.
\label{finsingeq}
\ee
It is easy to see that the l.h.s. of this equation is invariant under $SL(2,\IZ)$ transformations \eqref{Sdualtr}.
This is in perfect agreement with the expectation that the full HM metric \eqref{mett1} must be S-duality invariant.

Now we are in a position to infer about the fate of the singularity in the presence of D(-1)-instantons.
To this end, let us set $\tau_1=0$, apply S-duality transformation
\be
\tau_2\to \tau_2^{-1},
\qquad
V\to V\tau_2^3
\label{Str}
\ee
in the singularity equation, and extract the limit $\tau_2\to 0$.
To accomplish the last step, it is convenient to work with equation \eqref{singeq-red} where the resummation has not been done yet.
The point is that after the transformation \eqref{Str} the limit of small $\tau_2$ is similar to the weak string coupling limit
before the transformation. In particular, all instanton contributions are exponentially suppressed and can be dropped.
As a result, the expansion of the l.h.s. of the singularity equation reads
\be
-\frac{\zeta(3)^2\chi_\CYm^2}{2(16\pi^3 V)^2}\,\tau_2^{-6}
+\frac{\zeta(3)\chi_\CYm^2}{6(16\pi^2 V)^2}\, \tau_2^{-4}
+\frac{\zeta(3)\chi_\CYm}{32\pi^3 V}\,\tau_2^{-3}+  \frac{\chi_\CYm}{96\pi V}\,\tau_2^{-1}
+1+O\bigl(e^{-2\pi\tau_2^{-1}}\bigr).
\ee
We see that the dominant term comes from the last term in \eqref{singeq-red}
due to the presence of the volume factor, which after the transformation \eqref{Str} generates the additional factor $\tau_2^{-6}$.
Its crucial feature is that it comes with the minus sign. Thus, we conclude that in the region of small $\tau$ the l.h.s of the singularity equation
is negative. This implies that the equation always has a solution and the inclusion of D(-1)-instantons is not sufficient to resolve the singularity.
Moreover, the situation in a sense becomes even worse because this conclusion was achieved independently on the sign
of the Euler characteristic and therefore, in contrast to the case of the perturbative metric,
the singularity appears now for both signs of $\chi_\CY$!

We do not expect that inclusion of D1-instantons or even D5 and D3-instantons will improve the situation.
It seems that the singularity can be resolved only in the full non-perturbative metric which, in particular,
takes into account the effects from NS5-brane instantons. Since they scale like $e^{-2\pi V\tau_2^2}$ \cite{Becker:1995kb},
they become dominating at strong coupling and can significantly change the behavior of the metric.

%--------------------------------------------------------------------------------------
\section{Discussion}
\label{sec-disc}

In this paper we computed an explicit expression for the metric on the HM moduli space of type II string theory compactified on
a Calabi-Yau threefold affected by D-instantons. In fact, we were not able to get the {\it exact} \qk metric which includes them all.
Instead, our result applies in two cases:
\begin{itemize}
\item
One includes {\it all} D-instantons, but the metric is not valid beyond the one-instanton approximation. In particular,
it is only {\it approximately} \qk.
\item
One includes only ``a half" of D-instantons by restricting to a set of charges which satisfy
the condition of mutual locality \eqref{mutloc}, and in fact can always be rotated to have vanishing magnetic components.
Then the metric is {\it exactly} \qk.
\end{itemize}
Actually, the set of electrically charged D-instantons is likely to be the maximal one for which it is possible to get
exact analytic expressions. Beyond this approximation, deriving the metric requires solution of a system of integral equations \eqref{eqXigi}.
These equations have the form of Thermodynamic Bethe Ansatz (TBA) \cite{Zamolodchikov:1989cf,Gaiotto:2008cd,Alexandrov:2010pp}
and, typically, it is impossible to solve the TBA equations analytically. Thus, at this point it is not evident whether our all orders result
can be further improved.

We also checked that in the four-dimensional case, our metric agrees with the Tod ansatz for QK metrics with one continuous isometry
and provides a function which is an {\it exact} non-trivial solution of the Toda equation.
The latter is defined implicitly by the two equations \eqref{dilUHM}.

Finally, we investigated the effect of D-instantons on the curvature singularity appearing in the one-loop corrected metric.
In particular, we found the equation determining the singularity hypersurface inside the moduli space and explicitly demonstrated,
restricting to the sector of D(-1)-instantons
in type IIB string theory, that it is invariant under $SL(2,\IZ)$-transformations.
Using this property, which allows to relate the weak and strong couplings, we showed that the singularity is not resolved.
We expect that the resolution will be possible only after taking into account contributions of NS5-brane instantons,
which are believed to cure some other problems of the D-instanton corrected HM moduli space as well.
For instance, in \cite{Pioline:2009ia} it was argued that they should regularize the divergence which occurs
in summing D-instantons over the charge lattice due to the exponential growth of DT invariants.
Although recently some progress has been achieved in formulating these non-perturbative effects in the twistorial framework
\cite{Alexandrov:2010ca,Alexandrov:2014mfa,Alexandrov:2014rca}, neither of the above issues has been addressed yet.

Regarding possible applications of our results, we would like to mention that the metric \eqref{mett1}
possesses several continuous isometries, which can be used to get a gauged supergravity in four dimensions.
The latter has a non-trivial scalar potential, which depends on the gauged isometries and on the metrics on both moduli spaces
of vector and hypermultiplets \cite{deWit:2001bk}. Although this potential was extensively studied for the gaugings
which start from the tree level c-map metric on $\cM_H$ (see, for example, the recent exhaustive work \cite{Fre:2014pca}),
there were just few attempts to incorporate the instanton effects in it by replacing the c-map metric with
the instanton corrected one \cite{Ketov:2002fu,Ketov:2014roa}. Our result appears as a natural starting point for such investigation.

%----------------------------------------------------------------------------------------
\section*{Acknowledgments}
%----------------------------------------------------------------------------------------
We are grateful to Sergei Ketov, Boris Pioline, Frank Saueressig, Ulrich Theis and Stefan Vandoren for valuable discussions.

\appendix

%-----------------------------------------------------------------------------------------
\section{Deriving the metric from twistor data}
\label{apap-metricgen}

The general procedure to derive the metric on a QK manifold $\cM$ from the knowledge of Darboux coordinates
on its twistor space $\cZ$ was described in detail in \cite{Alexandrov:2008nk} and briefly outlined in section \ref{subsec-Dinst}.
Here we recapitulate the main steps and provide all relevant equations.

The starting point is the Laurent expansion of the Darboux coordinates near $t=0$.
We assume that it is given by
\be
\begin{split}
\xii{+}^\Lambda=&\,\xii{+}^{\Lambda,-1}\varpi^{-1}+\xii{+}^{\Lambda,0}+O(\varpi),
\\
\txii{+}_\Lambda=&\,\txii{+}_{\Lambda,0}+O(\varpi),
\\
\ai{+}=&\,4\I c\log \varpi+\ai{+}_{0}+O(\varpi),
\end{split}
\label{expDc}
\ee
where the index $\scriptstyle{[+]}$ indicates the patch surrounding the north pole of $\CP$.
This assumption is consistent with the form of Darboux coordinates in the case of the D-instanton corrected HM moduli space.
Next, one should proceed with the following four steps:
\begin{itemize}
\item
Substituting the expansions \eqref{expDc} into the contact one-form $\cX$ \eqref{contform}
and comparing it with the canonical form $Dt$ \eqref{canform} using \eqref{relcontform}, one finds the components of
the SU(2) connection
\be
\begin{split}
p^+ &=\frac{\I}{4}\, e^{-\phi}
\, \xi^{\Lambda,-1}_{[+]}  \de\txi^{[+]}_{\Lambda,0},
\\
p^3 &= -\frac{1}{4}\, e^{-\phi} \left( \de\alpha^{[+]}_0 +
\xi^{\Lambda,0}_{[+]}  \de\txi^{[+]}_{\Lambda,0} +
\xi^{\Lambda,-1}_{[+]}  \de\txi^{[+]}_{\Lambda,1}  \right) .
\end{split}
\label{connection}
\ee

\item
Then one computes the triplet of quaternionic 2-forms \eqref{su2curvature}.
In particular, for $\omega^3$ the formula reads
\be
\omega^3 = -2{\rm d} p^3+ 4\I  p^+ \wedge p^-.
\label{Kform}
\ee

\item
One specifies the almost complex structure $J^3$ by providing a basis of (1,0) forms on $\cM$.
Such a basis was found in \cite{Alexandrov:2008nk} and, after some simplifications, it takes the following form
\be
\label{defPi}
\pi^a =\de \(\xii{+}^{a,-1}/\xii{+}^{0,-1}\) ,
\qquad
\tilde\pi_\Lambda= \de\txii{+}_{\Lambda,0}  ,
\qquad
\tilde\pi_\alpha = \frac{1}{2\I}\,\de\ai{+}_0 +2c \,\de\log\xii{+}^{0,-1}.
\ee

\item
Finally, the metric is recovered as $g(X,Y) = \omega^3(X,J^3 Y)$.
To do this in practice, one should rewrite $\omega^3$, computed by \eqref{Kform} in terms of differentials of (generically real)
coordinates on $\cM$, in the form which makes explicit that it is of (1,1) Dolbeault type.
Using for this purpose the basis $\pi^X=(\pi^a,\tilde\pi_\Lambda,\tilde\pi_\alpha)$ given in \eqref{defPi}, the final result should look like
\be
\omega^3=\I g_{X\bY} \pi^X\wedge \bar\pi^{Y},
\label{metom}
\ee
from which the metric readily follows as $\de s^2 =2 g_{X\bY} \pi^X \otimes \bar\pi^{Y}$.

\end{itemize}

%-------------------------------------------------------------------------------------------
\section{Details of the metric evaluation}
\label{ap-metric}

In this appendix we provide some technical details on the derivation of the metric \eqref{mett1} and its properties.
First, in section \ref{notations} we collect various useful notations and relations. Section \ref{ap-om3first}
provides intermediate results leading the expression for the quaternionic 2-form $\omega^3$ in the coordinate basis.
Next section is devoted to rewriting this 2-form in the basis of (1,0)-forms.
Finally, in section \ref{ap-sympl} we check the symplectic invariance of the resulting metric.

\subsection{Notations and useful relations}
\label{notations}

In course of the presentation we have repeatedly used several convenient notations to make the equations look concise.
In this subsection we provide a list of these notations and some useful properties.
This should facilitate the reader for looking them up and helps us to avoid introducing them in scattered fashion throughout the text.
To better distinguish different sets of definitions, we put them under separate items.

\begin{itemize}
\item

First, we introduce functions on the moduli space, which can all be obtained as Fourier coefficients of $\Igg(\varpi)$ or
its derivative with respect to one of the moduli, around $t=0$ and $t=\infty$,
\be
\begin{array}{rclrcl}
\Igg{}& = &\displaystyle
\int_{\ellg{\gamma}}\frac{\d \varpi}{\varpi}\,
\log\(1-\qr e^{-2\pi \I \Xi_\gamma(\varpi)}\),
\quad &
\rIg&=&
\displaystyle \int_{\ellg{\gamma}}\frac{\d \varpi}{\varpi}\,
\frac{1}{\qr e^{2\pi \I \Xi_\gamma(\varpi)}-1},
\\
\Igpm& = &\displaystyle
\pm\int_{\ellg{\gamma}}\frac{\d \varpi}{\varpi^{1\pm 1}}\,\log\(1-\qr e^{-2\pi \I \Xi_\gamma(\varpi)}\),
\quad &
\rIgpm&=&
\displaystyle \pm \int_{\ellg{\gamma}}\frac{\d \varpi}{\varpi^{1\pm 1}}\,
\frac{1}{\qr e^{2\pi \I \Xi_\gamma(\varpi)}-1},
\end{array}
\label{newfun-expand}
\ee
where $\Xi_\gamma(\varpi)$ is given in \eqref{eqXigi-ml}.
They satisfy the reality properties
\be
\overline{\Ingam{n}{\gamma}}=\Ingam{n}{-\gamma},
\qquad
\overline{\Insgam{n}{+}{\gamma}}=\Insgam{n}{-}{-\gamma}
\ee
and the following identities
\be
\Zg{}\Insgam{n}{+}{\gamma}=\bZg{}\Insgam{n}{-}{\gamma},
\ee
which can be established by partial integration.

\item
Next, we introduce a useful shorthand notation
\be
V_{\gamma\Lambda}= q_\Lambda -F_{\Lambda\Sigma}p^\Sigma,
\label{defVgam}
\ee
and a function which depends on two charges
\be
\cQ_{\gamma\gamma'}= N^{\Lambda\Sigma}\Re V_{\gamma\Lambda}\Re V_{\gamma'\Sigma}+ \frac14\, N_{\Lambda\Sigma} p^\Lambda p'^\Sigma.
\label{defmatQ}
\ee
We would like to consider it as a matrix acting on the (infinite-dimensional) space of vectors whose components are enumerated by charges.
The above introduced $V_{\gamma\Lambda}$ is an example of such vector.
Note that for mutually local charges, the matrix $\cQ_{\gamma\gamma'}$ satisfies the following useful relations,
\be
\cQ_{\gamma\gamma'}= V_{\gamma\Lambda} N^{\Lambda\Sigma}\bV_{\gamma'\Sigma}
=\bV_{\gamma\Lambda}N^{\Lambda\Sigma} V_{\gamma'\Sigma}.
\label{QV}
\ee

\item
Then, we define two matrices which play a very important role in our story.
One of them is an instanton corrected version of $N_{\Lambda\Sigma}$ and the other is a matrix
on the space of charge labeled vectors as above,
\bea
\Min_{\Lambda\Sigma}&=&  N_{\Lambda\Sigma}-\frac{1}{2\pi}\sum_\gamma \Om{\gamma}\rIg \bV_{\gamma\Lambda}V_{\gamma\Sigma},
\label{defmatM}
\\
\Min_{\gamma\gamma'}&=&
\delta_{\gamma\gamma'}-\frac{\Om{\gamma'}}{2\pi}\,\rIgam{\gamma}\cQ_{\gamma\gamma'}.
\label{defmatA}
\eea
These two matrices are not independent of each other. The property \eqref{QV} ensures that their inverse matrices satisfy the following
relation, which allows to express $\Min_{\gamma\gamma'}^{-1}$ in terms of $(\Min^{-1})^{\Lambda\Sigma}\equiv\Min^{\Lambda\Sigma}$,
\be
\Min^{-1}_{\gamma\gamma'}=\delta_{\gamma\gamma'}+\frac{1}{2\pi}\,\rIg V_{\gamma\Lambda}\Min^{\Lambda\Sigma}\bV_{\gamma'\Sigma}\Om{\gamma'}.
\label{relMinv}
\ee

\item
It is convenient also to introduce several vectors
\be
\begin{split}
\vv_\gamma=&\,\frac{1}{4\pi}\( Z_\gamma\rIgp+\bZ_\gamma\rIgm\),
\\
\vvb_\gamma=&\,\sum_{\gamma'}\Om{\gamma}\Om{\gamma'}\cQ_{\gamma\gamma'}\vv_{\gamma'},
\\
\vl_\Lambda=&\,\sum_\gamma \Om{\gamma} \vv_\gamma V_{\gamma\Lambda},
\end{split}
\label{notvec}
\ee
a potential
\be
\Uk=K-\frac{1}{2\pi}\sum_\gamma \hng{}|Z_\gamma|^2 \rIg,
\label{defUpot}
\ee
and another potential and a vector labeled by charges, which have two representations due to the relation \eqref{relMinv},
\bea
\Uin&=&\Uk+\sum_{\gamma,\gamma'} \vvb_\gamma\Min^{-1}_{\gamma\gamma'}\vv_{\gamma'}
\nn\\
&=&K-\frac{1}{2\pi}\sum_\gamma \hng{}|Z_\gamma|^2 \rIg+ \vl_\Lambda \Min^{\Lambda\Sigma}\bvl_\Sigma.
\label{defUin}
\\
\cW_\gamma &=&  \bZg{}\rIg-\hng^{-1}\rIgp\sum_{\gamma'}\vvb_{\gamma'} \Min^{-1}_{\gamma'\gamma}
\nn\\
&=&\bZ_\gamma \rIg- \rIgp \vl_\Lambda\Min^{\Lambda\Sigma} \bV_{\gamma\Sigma}.
\label{cWdef}
\eea

\item
Finally, we define two 1-forms. The first one is a certain linear combination of the differentials of the RR-fields
\be
\cCf_\gamma= N^{\Lambda\Sigma}\(q_\Lambda-\Re F_{\Lambda\Xi}p^\Xi\)\(\de\tzeta_\Sigma-\Re F_{\Sigma\Theta}\de\zeta^\Theta\)
+\frac14\, N_{\Lambda\Sigma}\,p^\Lambda\,\de\zeta^\Sigma,
\label{connC}
\ee
which is built in the way analogous to $\cQ_{\gamma\gamma'}$ \eqref{defmatQ}.
The second, which we call $\cV$, appears explicitly in the HM metric \eqref{mett1} and arises as
the imaginary part of a certain (1,0)-form, see \eqref{ImSig} below.
In terms of $\cCf_\gamma$ and the other quantities introduced above, it reads
\bea
\cV&=&
2\cR^2 K\(1-\frac{4r}{\cR^2\Uin}\) \cA_K
+\frac{8r}{\cR\Uin}\sum_\gamma\hng{} \( \vv_\gamma+\frac{1}{2\pi}\, \rIg V_{\gamma\Lambda}\Min^{\Lambda\Sigma}\bvl_\Sigma\)\cCf_\gamma
\nn\\
&& +\frac{2r }{\pi\I\Uin}\sum_\gamma \Om{\gamma} \[\(\cW_\gamma+\frac{\cR\Uin}{8\pi\I r}\,\Igp\)\de Z_\gamma
-\(\bar\cW_\gamma+\frac{\cR\Uin}{8\pi\I r}\,\Igm\)\de \bZ_\gamma  \].
\label{conn}
\eea

\end{itemize}

\subsection{Computation of $\omega^3$}
\label{ap-om3first}

The coefficients of the Laurent expansion of the Darboux coordinates:
\begin{subequations}
\bea
\xii{+}^{\Lambda,-1} &=& \cR z^\Lambda,
\\
\xii{+}^{\Lambda,0}&=&\zeta^\Lambda-\frac{1}{8\pi^2}\sum\limits_{\gamma} \hng{} p^\Lambda \Igg{},
\\
\txii{+}_{\Lambda,0}&=&\tzeta_\Lambda-F_{\Lambda\Sigma}\zeta^\Sigma
-\frac{1}{8\pi^2}\sum\limits_{\gamma} \hng{} V_{\gamma\Lambda}\Igg{},
\label{Lexptxi}
\\
\txii{+}_{\Lambda,1}&=&-\I\cR\bz^\Sigma N_{\Lambda\Sigma}
-\frac{1}{2\cR}\, F_{\Lambda\Sigma\Theta}\zeta^\Sigma\zeta^\Theta
-\frac{1}{4\pi^2}\sum\limits_{\gamma} \hng{} \bigg[V_{\gamma\Lambda}\Igamp{\gamma}
\nn\\
&&
-\frac{1}{2\cR}\, F_{\Lambda\Sigma\Theta}p^\Sigma\zeta^\Theta\Igg{}
+\frac{1}{32\pi^2\cR}\, F_{\Lambda\Sigma\Theta}p^\Sigma\sum_{\gamma'} \Om{\gamma'} p'^\Theta \Igg{}\Igam{\gamma'}\bigg],
\\
\alpi{+}_{0}&=&
-\hf\(\sigma+\zeta^\Lambda\tzeta_\Lambda-F_{\Lambda\Sigma}\zeta^\Lambda\zeta^\Sigma\)+2\I\(r+c\)
-\frac{1}{8\pi^2}\sum\limits_{\gamma} \hng{} \[
\frac{1}{2\pi\I }\int_{\ellg{\gamma}}\frac{\d \varpi'}{\varpi'}\,
\Li_2\(\qr e^{-2\pi \I \Xigi{}(\varpi')}\)
\right.
\nn\\
&& \left.
- V_{\gamma\Lambda}\zeta^\Lambda \Igg{}
-\cR\Zg{}\Igp
+\frac{1}{16\pi^2}\, p^\Lambda\Igg{}\sum_{\gamma'}\Om{\gamma'} V_{\gamma'\Lambda}\Igam{\gamma'}\].
\eea
\label{LaurentDarboux}
\end{subequations}
The components of the SU(2) connection:
\be
\begin{split}
p^+ =&\,\frac{\I}{4r}
\left[ \cR z^\Lambda\(\de\tzeta_\Lambda-F_{\Lambda\Sigma}\de\zeta^\Sigma\)
-\frac{\cR}{8\pi^2}\sum\limits_{\gamma} \hng{} \Zg{}\de \Igg{}
\right] ,
\\
p^3 =&\, \frac{1}{8r} \left[ \de\sigma
+\tzeta_\Lambda\de \zeta^\Lambda-\zeta^\Lambda\de \tzeta_\Lambda
+2\cR^2 K\cA_K
-\frac{\cR}{4\pi^2}\sum\limits_{\gamma} \hng{} \(\Igp \de\Zg{}-\Igm \de\bZg{}\) \right].
\end{split}
\label{SU(2)connect}
\ee
The quaternionic 2-form:
\bea
\omega^3 &= & \frac{1}{4 r^2}\, \de r\wedge \[ \de\sigma
+\tzeta_\Lambda\de \zeta^\Lambda-\zeta^\Lambda\de \tzeta_\Lambda
-\frac{\cR}{4\pi^2}\sum\limits_{\gamma} \hng{} \(\Igp \de\Zg{}-\Igm \de\bZg{}\) \]
\nn\\
&&
+\frac{\cR^2 K}{2 r}\, \de\log\frac{ r}{\cR^2}\wedge\cA_K
+\frac{1}{2r}\(\de\zeta^\Lambda\wedge \de\tzeta_\Lambda
-\I\cR^2 N_{\Lambda\Sigma}\de z^\Lambda \wedge\de\bz^\Sigma
+\frac{\I\cR^2}{2r}\,z^\Lambda\bz^\Sigma\cY_\Lambda\wedge\bar\cY_\Sigma\)
\nn\\
&& +\frac{1}{16\pi^2 r}\sum_\gamma\hng{}\Bigl(\de \Igp \wedge\de\(\cR\Zg{}\)-\de\Igm\wedge \de\(\cR \bZg{}\) \Bigr).
\label{om3}
\eea

\subsection{$\omega^3$ in the holomorphic basis}
\label{ap-om3}

In this subsection we reexpress the quaternionic 2-form $\omega^3$ \eqref{om3} in the basis of (1,0)-forms given explicitly in
the main text, see \eqref{holforms}.
To this end, we start from the terms involving $\de \sigma$. There is only one such term in \eqref{om3} and, similarly, in \eqref{holforms}
it appears only in $\Im\Sigma$. Therefore, to rewrite the contribution $\de r\wedge \de \sigma$ as a part of a 2-form which is manifestly of type (1,1),
it is natural to look for the combination
\be
2\I \hat\Sigma\wedge \bar{\hat\Sigma}=4\Re\hat\Sigma\wedge \Im\hat\Sigma,
\label{prodSigma}
\ee
where $\hat\Sigma$ is a (1,0)-form
\be
\hat\Sigma=\Sigma+f_\Lambda\de z^\Lambda+g^\Lambda \cY_\Lambda,
\label{hatsigma}
\ee
fixed by the requirement that
\be
\Re\hat\Sigma\sim \de r.
\label{constr1}
\ee
Substituting the explicit expressions for (1,0)-forms into this condition, expressing $\de\log\cR$ in terms of $\de r$ and differentials
of other coordinates via the relation obtained by taking differential of \eqref{dil}
\be
\de r= \frac{\cR^2}{2}\,\Uk\, \de \log\cR+\frac{\cR^2}{4}\, \de K
+\frac{\cR}{4}\sum_\gamma \hng{}\(\vv_\gamma \de \Theta_\gamma-\frac{\cR}{2\pi}\,\rIg\de |\bZ_\gamma|^2\),
\label{res-dr}
\ee
where we used notations from \eqref{notvec} and \eqref{defUpot}, and equating to zero the coefficients of all one-forms except $\de r$,
one arrives at the following two equations on $f_\Lambda$ and $g^\Lambda$
\bea
f_\Lambda&=&\hf\, \cR^2\(1-\Ak(g)\) K_\Lambda
+\frac{\cR}{4\pi}\sum_\gamma\hng{} V_{\gamma\Lambda}
\biggl[\frac{\Igp}{2\pi\I}+\cR\Ak(g)\bZg{}\rIg
\nn  \\
&& \qquad\qquad\qquad \qquad\qquad
-2\Im\(g^\Sigma V_{\gamma\Sigma}\)\rIgp\biggr],
\label{constrZ}
\\
0&=& \Re g^\Lambda\(\de\tzeta_\Lambda-\Re F_{\Lambda\Sigma}\de\zeta^\Sigma\)+\Im g^\Lambda \Im F_{\Lambda\Sigma}\,\de\zeta^\Sigma
\nonumber\\
&&+\frac{1}{4}\sum_\gamma\hng{} \[\cR\Ak(g)\vv_\gamma +\frac{1}{\pi}\,\Im\(g^\Sigma V_{\gamma\Sigma}\)\rIg
\]\de\Theta_\gamma,
\label{constrTheta}
\eea
where we introduced the function of $g^\Lambda$
\be
\Ak(g)=\frac{4}{\cR^2\Uk}\(r-\frac{\cR}{2}\sum_\gamma \hng{}\Im\(g^\Sigma V_{\gamma\Sigma}\)\vv_\gamma\).
\label{deflam}
\ee
To solve the second condition \eqref{constrTheta} with respect to $g^\Lambda$, one can use the following trick.
Let us choose the ansatz
\bea
\Re g^\Lambda =- \sum_\gamma \hng{} p^\Lambda g_\gamma,
\qquad
\Im g^\Lambda =-2  \sum_\gamma \hng{} N^{\Lambda\Sigma}\(q_\Sigma-\Re F_{\Sigma\Xi}p^\Xi\)g_\gamma,
\label{eqg}
\eea
where $g_\gamma$ is still to be found.
Then all terms in \eqref{constrTheta} become proportional to $\de\Theta_\gamma$
and the condition reduces to a linear equation on $g_\gamma$,
\be
g_\gamma=-\frac{1}{4}\,\cR\Ak(g) \vv_\gamma
+\frac{1}{2\pi}\,\rIg\sum_{\gamma'}\Om{\gamma'}\cQ_{\gamma\gamma'}g_{\gamma'},
\label{eqgg}
\ee
where the last term is written using the matrix \eqref{defmatQ}.
As a result, the solution for $g_\gamma$ is obtained as
\be
g_\gamma =-\frac{r}{\cR\,\Uk}\sum_{\gamma'}\Minst^{-1}_{\gamma\gamma'}\vv_{\gamma'},
\label{solggam}
\ee
where $\Minst^{-1}_{\gamma\gamma'}$ is the inverse of the matrix
\be
\Minst_{\gamma\gamma'}
=\Min_{\gamma\gamma'}+\Uk^{-1}\vv_\gamma \vvb_{\gamma'}
\label{Minstcharge}
\ee
and we used another matrix introduced in \eqref{defmatA} and the vectors from \eqref{notvec}.
Since the last term is just the product of two vectors, $\Minst^{-1}_{\gamma\gamma'}$ can be expressed in terms of the inverse of
$\Min_{\gamma\gamma'}$. An easy calculation shows that
\be
\Minst^{-1}_{\gamma\gamma'}=
\sum_{\gamma''}\Min^{-1}_{\gamma\gamma''}\[\delta_{\gamma''\gamma'}
-\frac{\sum_{\tlgam}\vvb_{\tlgam}\Min^{-1}_{\tlgam\gamma'}}{\Uk+\sum_{\tlgam,\tlgam'}\vvb_{\tlgam}\Min^{-1}_{\tlgam\tlgam'}\vv_{\tlgam'}}
\, \vv_{\gamma''}\],
\label{proprty-mat}
\ee
which leads to the following simple result
\be
g_{\gamma}= -\frac{r}{\cR\,\Uin}\sum_{\gamma'}\Min^{-1}_{\gamma\gamma'}\vv_{\gamma'},
\ee
where comparing to \eqref{solggam} the potential $\Uk$ has been converted into $\Uin$ defined in \eqref{defUin}.
Substituting this result into \eqref{deflam} and \eqref{constrZ}, one finds
\bea
\Ak(g)&=&\frac{4r}{\cR^2\Uin},
\\
f_\Lambda &=&\(\frac{\cR^2}{2}-\frac{2r}{\Uin}\)\, N_{\Lambda\Sigma}\bz^\Sigma
+\frac{1}{\pi}\sum_\gamma\hng{} V_{\gamma\Lambda}
\( \frac{\cR}{8\pi\I}\, \Igp +\frac{r}{\Uin}\, \cW_\gamma\),
\eea
where we used\footnote{To get this quantity, we used the property that the matrix
$$
\cS_{\gamma\gamma'}=\Om{\gamma}\sum_{\gamma''}\Om{\gamma''}\cQ_{\gamma\gamma''}\Min_{\gamma''\gamma'}^{-1}
$$
is symmetric. Its symmetricity is equivalent to the symmetricity of
$$
\rIg\cS_{\gamma\gamma'}\rIgam{\gamma'}
=-2\pi \Om{\gamma}\sum_{\gamma''}\(\delta_{\gamma\gamma''}-\Min_{\gamma\gamma''}\)\Min_{\gamma''\gamma'}^{-1}\rIgam{\gamma'}
=-2\pi \Om{\gamma}\(\Min_{\gamma\gamma'}^{-1} -\delta_{\gamma\gamma'}\)\rIgam{\gamma'}
$$
which is indeed symmetric. \label{footS}}
$\cW_\gamma$ defined in \eqref{cWdef}.
Having found these solutions, it is now straightforward to check that the condition \eqref{constr1}
is indeed satisfied. This follows from
\bea
\Re\hat\Sigma&=& 2 \(1-\frac{2r}{\cR^2\Uin}\)\de r ,
\label{reSig}
\\
\Im\hat\Sigma&=& \frac{1}{4}\(\de \sigma+\tzeta_\Lambda\de \zeta^\Lambda-\zeta^\Lambda\de \tzeta_\Lambda+\cV\) ,
\label{ImSig}
\eea
where the one-form $\cV$ is explicitly given in \eqref{conn}.
As a result, for the only contribution containing $\de \sigma$ one finds
\be
\frac{1}{4 r^2}\, \de r\wedge \(\de\sigma+\tzeta_\Lambda\de \zeta^\Lambda-\zeta^\Lambda\de \tzeta_\Lambda\)
 =\frac{\I \hat\Sigma\wedge \bar{\hat\Sigma}}{4 r^2\(1-\frac{2r}{\cR^2\Uin}\)}
-\frac{1}{16r^2}\, \de r\wedge \cV,
\ee
where the last term is independent of $\de \sigma$.
Substituting this into \eqref{om3}, one obtains the following intermediate expression for $\omega^3$
\bea
\omega^3 &= &  \frac{\I\,\hat\Sigma\wedge \bar{\hat\Sigma}}{4 r^2\(1-\frac{2r}{\cR^2\Uin}\)}
+\frac{\I\cR^2}{4r^2}\,z^\Lambda\bz^\Sigma\cY_\Lambda\wedge\bar\cY_\Sigma
-\frac{\I\cR^2}{2r}\, N_{\Lambda\Sigma}\de z^\Lambda \wedge\de\bz^\Sigma
\nn\\
&&
+\frac{1}{2r}\,\de\zeta^\Lambda\wedge \de\tzeta_\Lambda
+\frac{1}{16\pi^2 r}\sum_\gamma\hng{}\(\de \Igp \wedge\de(\cR\Zg{})-\de\Igm\wedge \de(\cR\bZg{}) \)
\label{inter-om3}\\
&& -\frac{\cR^2 K}{ r }\,\de\log\cR\wedge\cA_K
+\frac{2\de r}{ r^2}\wedge
\biggl[  \frac{r K}{\Uin}\,\cA_K +\sum_\gamma\hng{} \(g_\gamma\cCf_\gamma
-\frac{r}{4\pi\I\Uin}\(\cW_\gamma \de Z_\gamma-\bar\cW_{\gamma}\de \bZ_\gamma\)\)\biggr].
\nn
\eea

The next terms to consider are those which are quadratic in the differentials of the RR-fields.
There are three such terms in \eqref{inter-om3}: the second term, which is already in the desired form;
the forth term; and one more contribution comes from the last term after substitution of \eqref{res-dr}.
To deal with them, we introduce a convenient notation
\be
\cY_\gamma=\I N^{\Lambda\Sigma}\, \bV_{\gamma\Lambda}\, \cY_\Sigma
\label{defUgam}
\ee
and note the following identity
\be
\begin{split}
&
\de\zeta^\Lambda\wedge\de\tzeta_\Lambda
+\frac{\cR}{r}\sum_{\gamma}\Om{\gamma}\vv_{\gamma}\de\Theta_{\gamma}\wedge \sum_{\gamma'}\Om{\gamma'}g_{\gamma'}\cCf_{\gamma'}
\\
=&\,
-\I N^{\Lambda\Sigma}\cY_\Lambda\wedge\bar\cY_\Sigma
+\I\sum_{\gamma,\gamma'}\(\frac{\cR^2\Uin}{r^2}\,  \Om{\gamma} \Om{\gamma'} g_\gamma g_{\gamma'}
-\frac{1}{2\pi}\,\Om{\gamma} \Min^{-1}_{\gamma\gamma'}\rIgam{\gamma'}\)\cY_\gamma\wedge \bar\cY_{\gamma'}
\\
&\, -\frac{1}{2\pi}\sum_{\gamma,\gamma'} \hng{}
\(\Min_{\gamma\gamma'}^{-1}+\frac{4 g_\gamma}{r}\sum_{\gamma''}\vvb_{\gamma''}\Min_{\gamma''\gamma'}^{-1}\)
\cCf_\gamma \wedge\Bigl(\rIgamp{\gamma'}\de (\cR Z_{\gamma'})+\rIgamm{\gamma'}\de(\cR\bZ_{\gamma'})\Bigr).
\end{split}
\ee
The l.h.s. represents exactly the contributions we wanted to rewrite. Thus, using this identity in \eqref{inter-om3}, one
puts $\omega^3$ in the form which is written using solely the (1,0)-forms \eqref{holforms}, their complex conjugate and $\de\cR$.
Furthermore, one can show that all terms involving $\de\cR$ cancel, as do also the terms of (2,0) and (0,2)-type.
After few manipulations, one can arrive at the following result
\bea
\omega^3 &= &  \frac{\I\,\hat\Sigma\wedge \bar{\hat\Sigma}}{4 r^2 \(1-\frac{2r}{\cR^2 \Uin}\)}
-\frac{\I}{2r}\(N^{\Lambda\Sigma}-\frac{\cR^2}{2r}\,z^\Lambda\bz^\Sigma\)\cY_\Lambda\wedge\bar\cY_\Sigma
-\frac{\I}{4\pi r}\sum_{\gamma,\gamma'}\Om{\gamma} \Min^{-1}_{\gamma\gamma'}\rIgam{\gamma'}\cY_\gamma\wedge \bar\cY_{\gamma'}
\nn\\
&&
+\frac{\I\cR^2}{8r^3\Uin}\sum_\gamma \Om{\gamma}\( \Uin g_\gamma\cY_\gamma+\frac{r}{\pi}\, \cW_\gamma\de Z_\gamma\)
\wedge \sum_{\gamma'}\Om{\gamma'}\(\Uin g_{\gamma'}\bar\cY_{\gamma'}+\frac{r}{\pi}\, \bar\cW_{\gamma'}\de \bZ_{\gamma'}\)
\nn\\
&&
+\frac{\I \cR}{4\pi r}\sum_{\gamma,\gamma'}\Om{\gamma}\Min^{-1}_{\gamma\gamma'}
\[\rIgamm{\gamma'}\cY_\gamma \wedge \(\de\bZ_{\gamma'}-\Uin^{-1}\bZ_{\gamma'}\bar\p K\)
+\rIgamp{\gamma'}\(\de Z_{\gamma'}-\Uin^{-1}Z_{\gamma'}\p K \)\wedge\bar\cY_\gamma\]
\nn\\
&&
+\frac{\I\cR^2}{2r}\[\Uin^{-1}\p K\wedge \bar \p K -N_{\Lambda\Sigma}\de z^\Lambda \wedge\de\bz^\Sigma
-\frac{1}{2\pi\Uin}\sum_{\gamma}\Om{\gamma}
\(\cW_{\gamma}\de Z_{\gamma}\wedge \bar\p K+\bar\cW_{\gamma}\p K\wedge \de \bZ_{\gamma}\)
\right.
\nn\\
&&\left.\qquad
+\frac{1}{2\pi}\sum_{\gamma,\gamma'} \(\hng{} \rIg\delta_{\gamma\gamma'} -
\frac{1}{2\pi}\cS_{\gamma\gamma'}\rIgp\rIgamm{\gamma'}\)\de Z_\gamma\wedge \de\bZ_{\gamma'}\],
\label{inter8-om3}
\eea
where in the last term the symmetric matrix $\cS_{\gamma\gamma'}$ was defined in footnote \ref{footS}.
The quaternionic 2-form \eqref{inter8-om3} is manifestly of (1,1)-type and the metric readily follows from it.

However, the result \eqref{inter8-om3} has one serious shortcoming: it requires to deal with infinite-dimensional
matrices and, in particular, to find the inverse $\Min^{-1}_{\gamma\gamma'}$. Fortunately, the situation can be improved
due to the relation \eqref{relMinv} that expresses $\Min^{-1}_{\gamma\gamma'}$ in terms of the inverse of another matrix,
which is already finie-dimensional.
In particular, this relation implies
\be
\begin{split}
\sum_{\gamma'} \vvb_{\gamma'}\Min^{-1}_{\gamma'\gamma}=&\, \Om{\gamma}\vl_\Lambda \Min^{\Lambda\Sigma} \bV_{\gamma\Sigma},
\\
\sum_{\gamma,\gamma'} \vvb_\gamma\Min^{-1}_{\gamma\gamma'}\vv_{\gamma'}
=&\,
\vl_\Lambda \Min^{\Lambda\Sigma}\bvl_\Sigma,
\\
\sum_\gamma \Om{\gamma} g_\gamma \cY_\gamma =&\, -\frac{\I r}{\cR\,\Uin}\,\cY_\Lambda \Min^{\Lambda\Sigma}\bvl_\Sigma,
\\
\sum_{\gamma} \Om{\gamma}\Min^{-1}_{\gamma\gamma'}\rIgam{\gamma'} \cY_\gamma
=&\, \I \Om{\gamma'}\rIgam{\gamma'}  \cY_\Lambda \Min^{\Lambda\Sigma}\bV_{\gamma'\Sigma},
\end{split}
\label{reltwoframes}
\ee
Using these identities, the 2-form \eqref{inter8-om3} can be rewritten
as follows
\bea
\omega^3&=&
\frac{\I \,\hat\Sigma\wedge \bar{\hat\Sigma}}{4\I r^2\(1-\frac{2r}{\cR^2\Uin}\)}
+\frac{\I\cR^2}{4r^2}\,z^\Lambda\bz^\Sigma \cY_\Lambda\wedge\bar\cY_\Sigma
\nn\\
&&
-\frac{\I}{2 r}\,\Min^{\Lambda\Sigma}\(\cY_\Lambda
+\frac{\I\cR}{2\pi}\sum_\gamma \Om{\gamma} V_{\gamma\Lambda}\rIgamp{\gamma}\(\de Z_{\gamma}-\Uin^{-1}Z_\gamma\p K\)\)
\nn\\
&&\qquad
\wedge \(\bar\cY_\Sigma-\frac{\I\cR}{2\pi}\sum_{\gamma'} \Om{\gamma'} \bV_{\gamma'\Sigma}
\rIgamm{\gamma'}\(\de \bZ_{\gamma'}-\Uin^{-1}\bZ_{\gamma'}\bar\p K\)\)
\nn\\
&&
+\frac{\I}{2r\Uin}\(  \cY_\Lambda \Min^{\Lambda\Sigma}\bvl_\Sigma-\frac{\I\cR}{2\pi}\, \sum_\gamma \Om{\gamma}\cW_\gamma\de Z_\gamma\)
\wedge\(\vl_{\Lambda'} \Min^{\Lambda'\Sigma'}\bar\cY_{\Sigma'}+\frac{\I\cR}{2\pi}\, \sum_{\gamma'} \Om{\gamma'}\bar\cW_{\gamma'}\de \bZ_{\gamma'}\)
\nn\\
&&
+\frac{\I\cR^2 K}{2r}\Biggl\{ \cK_{ab}\de z^a\wedge \de\bz^b
-\frac{1}{K^2\Uin^2}\(\frac{1}{2\pi}\sum_\gamma \hng{}|\Zg{}|^2 \rIg- \vl_\Lambda \Min^{\Lambda\Sigma}\bvl_\Sigma\)^2
\p K\wedge \bar \p K
\Biggr.
\nn\\
&&\Biggl.
+\frac{1}{2\pi K}\sum_\gamma \hng{} \rIg\(\de Z_\gamma-\Uin^{-1}Z_\gamma\p K\)\wedge \(\de\bZ_\gamma-\Uin^{-1}\bZ_\gamma\bar\p K\)
 \Biggr\}.
\label{final-om3}
\eea
The metric can be read off as $g(X,Y) = \omega^3(X,J^3 Y)$
and its explicit expression is presented in the main text, see \eqref{mett1}.
Finally, note that the (1,0)-form $\cY_\Lambda$ given in \eqref{holforms} can be expanded in the differentials of
the standard coordinates on the HM moduli space upon using \eqref{res-dr}. The result is given by
\be
\begin{split}
\cY_\Lambda
=&\,\de\tzeta_\Lambda-F_{\Lambda\Sigma}\de\zeta^\Sigma-\frac{\I}{4\pi}\sum_\gamma\hng{}V_{\gamma\Lambda}
\Biggl[ \rIg\de\Theta_\gamma +\cR\(\rIgp\de Z_\gamma+\rIgm\de\bZ_\gamma\)
\Biggr.
\\
&\, \Biggl.
\!\!\!\!+\frac{2\pi\cR}{\Uk}\,\vv_\gamma \(\frac{4\,\de r}{\cR^2}- \de K
-\sum_{\gamma'} \Om{\gamma'}\(\frac{\vv_{\gamma'}}{\cR}\, \de \Theta_{\gamma'}-
\frac{\rIgam{\gamma'}}{2\pi}\,\(\bZ_{\gamma'} \de Z_{\gamma'}+Z_{\gamma'}\de \bZ_{\gamma'}\)\)\)\Biggr].
\end{split}
\label{expand-Y}
\ee
An important feature of this result is that it shows that in the presence of instantons
$\cY_\Lambda$ has a non-vanishing projection along $\de r$.

\subsection{Check of symplectic invariance}
\label{ap-sympl}

In this appendix, we verify that the instanton corrected metric \eqref{mett1}, although it is not manifest,
is invariant under symplectic transformations. It turns out that the proof of the invariance
is much easier if one works in terms of infinite-dimensional matrix $\Min_{\gamma\gamma'}$ and not with
its finite-dimensional cousin $\Min_{\Lambda\Sigma}$. Due to this reason, we will consider the expression
\eqref{inter8-om3} for the quaternionic 2-form, rather than the expression for the metric presented in the main text.
Since the two are related by simple algebraic manipulations, symplectic invariance of one follows from that of the other.

The starting point is the matrix
\be
\cN_{\Lambda\Sigma}=\bF_{\Lambda\Sigma}-\I\,\frac{(Nz)_\Lambda (Nz)_\Sigma}{(zNz)}.
\ee
It plays an important physical role since it appears in the kinetic term for the gauge fields in the vector multiplet sector \cite{deWit:1984px}.
Its imaginary part is invertible and satisfies
\be
\hf\,\Im \cN^{\Lambda\Sigma}=N^{\Lambda\Sigma}-K^{-1}\(z^\Lambda\bz^\Sigma+\bz^\Lambda z^\Sigma\),
\ee
which shows that it also naturally arises in the kinetic term for the RR-fields at perturbative level, cf. the second term in \eqref{1lmetric}.
Its importance for our discussion follows from the fact that $\cN_{\Lambda\Sigma}$ and its imaginary part have nice
transformation properties under the symplectic group.
For an element
$\(\begin{array}{cc} A & B \\ C & D\end{array}\)\in Sp(2n,\IZ)$, they are given by \cite{deWit:1995jd}
\be
\begin{split}
\cN\ \mapsto&\ (C+D\cN)(A+B\cN)^{-1},
\\
\Im \cN\ \mapsto&\ (A+B\cN)^{-T}\Im \cN(A+B\bar\cN)^{-1}.
\end{split}
\label{transfN}
\ee
Besides, let us define for any symplectic vector $\rho=(\chi^\Lambda,\psi_\Lambda)$, which transforms in the defining representation of
$Sp(2n,\IZ)$, the two vectors of dimension $n$:
\be
V_{\rho\Lambda}=\psi_\Lambda-F_{\Lambda\Sigma}\chi^\Sigma,
\qquad
\cV_{\rho\Lambda}=\psi_\Lambda-\cN_{\Lambda\Sigma}\chi^\Sigma.
\ee
It is easy to check that the second vector defined by $\cN_{\Lambda\Sigma}$, in contrast to the first one, transforms as a modular form,
\be
\cV_\rho\ \mapsto\ (A+B\cN)^{-1}\cV_\rho.
\ee
These properties imply, in particular, that the combination $\cV_\rho\Im \cN^{-1} \bar\cV_{\tilde\rho}$ is symplectic invariant
for any symplectic vectors $\rho$ and $\tilde \rho$.

At the next step, let us again consider two symplectic vectors, $\rho$ and $\tilde\rho$, and define for them the following quantity
\be
\cQ(\rho,\tilde\rho)=\(\psi_\Lambda-\Re F_{\Lambda\Lambda'}\chi^{\Lambda'}\)N^{\Lambda\Sigma}
\(\tilde\psi_\Sigma-\Re F_{\Sigma\Sigma'}\tilde\chi^{\Sigma'}\)+\frac14\,\chi^\Lambda N_{\Lambda\Sigma}\tilde\chi^\Sigma.
\ee
It naturally appears in our context since, in particular, one has $\cQ_{\gamma\gamma'}=\cQ(\gamma,\gamma')$ and $\cCf_\gamma=\cQ(\gamma,\de C)$
where $C=(\zeta^\Lambda,\tzeta_\Lambda)$ is the vector of RR-fields. Then it is straightforward to prove that
\be
\begin{split}
\cQ(\rho,\tilde\rho)=&\, V_{\rho\Lambda} N^{\Lambda\Sigma}\bV_{\tilde\rho\Sigma}+\frac{\I}{2} \langl\rho,\tilde\rho\rangl
= \bV_{\rho\Lambda}N^{\Lambda\Sigma}V_{\tilde\rho\Sigma} - \frac{\I}{2} \langl\rho,\tilde\rho\rangl
\\
=&\, \frac12\,\cV_{\rho\Lambda}\Im \cN^{\Lambda\Sigma}\bar \cV_{\tilde\rho\Sigma}-\frac{\I}{2} \langl\rho,\tilde\rho\rangl
+\frac{1}{K}\(\langl \rho,X\rangl \langl \tilde\rho,\bX\rangl + \langl \rho,\bX\rangl \langl \tilde\rho,X\rangl\),
\end{split}
\ee
where $X=(z^\Lambda,F_\Lambda)$ and $\langle\,\cdot\,,\,\cdot\,\rangle$ is the symplectic invariant scalar product introduced
below \eqref{eqXigi}.
Since all terms on the r.h.s. are symplectic invariant, this result shows the invariance of
$\cQ(\rho,\tilde\rho)$ as well as of $V_\rho N^{-1}\bar V_{\tilde\rho}$.
In particular, the latter fact ensures the invariance of the perturbative metric \eqref{1lmetric}
because the only term, which is not manifestly symplectic invariant, is $V_{\de C}N^{-1} \bV_{\de C}$.

Now it is easy to prove the invariance of the instanton corrected HM metric. It is sufficient to note that
all non-manifestly invariant terms in $\omega^3$ \eqref{inter8-om3}
are constructed from the following building blocks:
$\cQ_{\gamma\gamma}$, $\cCf_\gamma$, $V_{\de C}N^{-1} \bV_{\gamma}$, and $V_{\de C}N^{-1} \bV_{\de C}$.
As we have just proved, they are symplectic invariant which, in particular, implies the invariance of
$\Uin$, $\cY_\gamma$, $\Min_{\gamma\gamma'}$ and the total metric.

For the purpose of rewriting the metric in a manifestly invariant form, it would be nice to promote $\Min_{\Lambda\Sigma}$
to a matrix which has transformation properties similar to $\Im\cN_{\Lambda\Sigma}$ \eqref{transfN}.
Unfortunately, this seems to be a hard task and we do not address it here.

%--------------------------------------------------------------------------------------------
\section{Match with the Tod ansatz}
%--------------------------------------------------------------------------------------------
\label{ap-Toda}

The aim of this appendix is to prove that in the case of the universal hypermultiplet the function $T=2\log (\cR/2)$
and the connection one-form $\cV$ satisfy the following two conditions
\bea
\p_r T&=&4\(\cR^2\Uin\)^{-1},
\label{firstcond}
\\
\de \cV&=& 2\de\zeta\wedge \de\tzeta+4\I r\(\p_z \p_r T\de z-\p_{\bz} \p_r T\de\bz\)\wedge \de r+16\I \p_r(P e^T)\de z\wedge \de \bz,
\label{secondcond}
\eea
and the Toda equation \eqref{eq-Toda}.

Specializing the notations from appendix \ref{notations} to the four-dimensional case and substituting
the prepotential \eqref{prepUHM}, one finds the following results\footnote{To get all these results as well as the ones which follow
below, it is crucial to take into account the condition of mutual locality, which takes the form $qp'=q'p$
and implies, in particular, that $Z_\gamma\bZ_{\gamma'}=\bZ_\gamma Z_{\gamma'}$ and $\vl\bZ_{\gamma}=\bvl Z_\gamma$.}
\be
\begin{split}
\vl =&\,\frac{1}{2\pi}\sum_\gamma \Om{\gamma} |Z_\gamma|^2 \rIgm ,
\\
\Min=&\, 2\lambda_2-\frac{1}{2\pi}\sum_\gamma \Om{\gamma} |Z_\gamma|^2 \rIg ,
\\
\Uin =&\, \Min+\Min^{-1}|\vl|^2,
\vphantom{A \over A}
\\
\cY =&\, \de\tzeta-\lambda \de\zeta-\frac{\I}{4\pi}\sum_\gamma \Om{\gamma}
Z_\gamma \(\rIg-\vl\Min^{-1}\rIgp\)\de \Theta_\gamma
-\frac{2\I\vl}{ \cR \Min}\, \de r,
\\
\cV=&\,\frac{4r}{\pi\cR\Uin}\sum_\gamma \Om{\gamma} Z_\gamma
\(\rIgp+\vl\Min^{-1} \rIg\)\cCf_\gamma.
\end{split}
\label{Ab-UHM}
\ee
Next, solving $\de z=\frac{\I}{2}\,\cY$ with respect to differentials of the RR-fields, one obtains
\be
\begin{split}
\de\zeta=&\, \frac{4}{\Uin}\Re\[ \frac{\vl}{\cR\Min}\, \de r
-\(1-\frac{\I}{4\pi}\sum_\gamma\hng p\( \rIg-\bvl\Min^{-1}\rIgm\)\bZ_\gamma\)\de z\],
\\
\de\tzeta=&\, \frac{4}{\Uin}\Re\[ \frac{\bar\lambda\vl}{\cR\Min}\, \de r
-\(\bar\lambda-\frac{\I}{4\pi}\sum_\gamma\hng q\( \rIg-\bvl\Min^{-1}\rIgm\)\bZ_\gamma\)\de z\].
\end{split}
\ee
Combining these differentials in various ways, one computes
\bea
\cCf_\gamma
&=& -\I  \, \(\bZ_\gamma \de z-Z_\gamma \de \bz\),
\label{combdzeta}
\\
\de\Theta_\gamma&=&-\frac{2}{\Uin}\( \bZ_\gamma \de z+ Z_\gamma \de \bz\)
+\frac{2(\vl\bZ_\gamma+\bvl Z_\gamma)}{\cR\Min\Uin}\, \de r.
\label{dTheta-UHM}
\\
\de\zeta\wedge \de\tzeta
&=&-\frac{4\I}{\Uin}\(\de z\wedge \de \bz-\frac{1}{\cR\Min}( \bvl\de z-\vl\de\bz)\wedge \de r\).
\label{dd-UHM}
\eea
Then differentiating \eqref{dil} and using \eqref{dTheta-UHM}, one gets
\be
\de \log\cR=\frac{2\de r}{\cR^2\Uin} +\frac{\bvl\de z+\vl\de\bz }{\cR\Min\Uin}.
\label{dR-UHM}
\ee
From this result, one immediately concludes that
\be
\p_r T=\frac{4}{\cR^2\Uin},
\qquad
\p_z T=\frac{2\bvl}{\cR\Min\Uin},
\label{diffsT}
\ee
which proves the first from our conditions \eqref{firstcond}.

To prove the second condition \eqref{secondcond}, we rewrite it in terms of differentials $\de r$ and $\de z$.
To this end, we use \eqref{dd-UHM} and \eqref{diffsT}. Then the r.h.s. of this condition can be put in the following form
\be
-8\I r \p_r^2 e^T\de z\wedge \de \bz
+4\I \Bigl[\(r\p_z \p_r T+\p_z T\)\de z-\(r\p_{\bz} \p_r T+\p_{\bz} T\)\de\bz\Bigr]\wedge \de r.
\label{rhsTh}
\ee
On the other hand, substituting \eqref{combdzeta} into the expression for the connection $\cV$ from \eqref{Ab-UHM},
one obtains a very simple result
\be
\begin{split}
\cV =&\,
=-4\I r \( \p_z T \de z -\p_{\bz} T\de\bz\).
\label{conn-UHM2}
\end{split}
\ee
It is trivial to see that its differential reproduces \eqref{rhsTh} provided $T$ satisfies the Toda equation \eqref{eq-Toda},
which proves the condition \eqref{secondcond}.

Finally, it remains to show that $T$ indeed fulfils the Toda equation.
Using the results for the first derivatives \eqref{diffsT}, it can be rewritten as
\be
\p_z\(\frac{2\vl}{\cR\Min\Uin}\)-\p_r\Uin^{-1}=0.
\ee
To demonstrate that this equation does hold, one then substitues explicit expressions for $\vl$, $\Min$ and $\Uin$ from \eqref{Ab-UHM}
and evaluates their derivatives. This is a straightforward, although a bit cumbersome exercise, and we prefer not to put it here.

%--------------------------------------------------------------------------------------------
\section{Poisson resummation}
%--------------------------------------------------------------------------------------------
\label{ap-resum}

In this appendix we perform the Poisson resummation of the functions $\BI,\CI$ \eqref{defS12} and $\AI$ \eqref{defS3}.
This procedure relies on the following resummation formula
\be
\sum_{q\in \IZ} f(q)=\sum_{n\in \IZ} g(2\pi n),
\qquad
g(w)=\int_{-\infty}^\infty \de x \,f(x)  e^{-\I w x}.
\ee

Before we start, it is convenient to note that all three functions can be written in terms of one family of functions
\be
f_m^{(\alpha)}(x)=(\sign x)^\alpha x^2 e^{-2\pi\I m x \zeta^0} \int_0^\infty \frac{\de s}{s}\( s^{-1}+s\)^\alpha e^{-2\pi m|x| \cR\( s^{-1}+s\)}.
\ee
Indeed, it is easy to check that
\be
\begin{split}
\BI =&\, \frac{\zeta(3)}{2\pi^2} -\hf\sum_{m>0}\sum_{q\in\IZ} f_m^{(2)}(q),
\\
\CI =&\, -\I \sum_{m>0}\sum_{q\in\IZ} f_m^{(1)}(q),
\\
\AI =&\, \frac{\zeta(3)}{2\pi^2}-\sum_{m>0}\sum_{q\in\IZ} f_m^{(0)}(q),
\end{split}
\label{ABCf}
\ee
where we used that $f_m^{(0)}(0)=f_m^{(1)}(0)=0$, whereas $f_m^{(2)}(0)=\frac{2}{(2\pi m\cR)^2}$.
This observation allows to perform the resummation in a uniform way because the only thing 
which should be evaluated is the Fourier transform of $f_m^{(\alpha)}(x)$.
Identifying $\zeta^0=\tau_1$ and $\cR=\tau_2/2$ according to the mirror map \eqref{mirmap} and
interchanging the order of integrations, one finds\footnote{The integral in \eqref{FT} converges only for $\alpha < 3$,
but this condition encompasses all the relevant cases for us.}
\bea
g_m^{(\alpha)}(2\pi n)
&=& \frac{1}{4\pi^3}\int_{-\infty}^\infty \frac{\de s}{s}\frac{\( \frac{1}{s}+s\)^\alpha }{\(\frac{m\tau_2}{2}\(s^{-1}+s\)+\I\(m\tau_1+n\)\)^3}
\nn\\
&=&\frac{\I}{4\pi^2}\(\frac{2}{m\tau_2}\)^3\left.\(\frac{\p^2}{\p s^2}\, \frac{s^2 \(s^{-1}+s\)^\alpha}{\(s+\I \tmmn\)^3}\)\right|_{s=-\I\tpmn},
\label{FT}
\eea
where $\tpmmn$ are the two roots of the denominator appearing in the first line, which are given explicitly by
\be
\tpmmn = \frac{ m \tau_1 + n \mp | m\tau + n |}{m \tau_2}.
\ee
Then a simple computation gives
\be
\begin{split}
g_m^{(0)}(2\pi n)=&\,  - \frac{1}{4\pi^2}\(\frac{2}{|m\tau+n|^3}-\frac{3(m\tau_2)^2}{|m\tau+n|^5}\),
\\
g_m^{(1)}(2\pi n)=&\,  - \frac{3\I}{2\pi^2}\,\frac{m\tau_2(m\tau_1+n)}{|m\tau+n|^5}\, ,
\\
g_m^{(2)}(2\pi n)=&\,  - \frac{1}{\pi^2}\(\frac{1}{|m\tau+n|^3}-\frac{3(m\tau_2)^2}{|m\tau+n|^5}\).
\end{split}
\ee
Applying the resummation formula in \eqref{ABCf} and plugging there these results, one immediately arrives at \eqref{AIBICI}.

\providecommand{\href}[2]{#2}\begingroup\raggedright\endgroup

%\bibliographystyle{utphys}
%\bibliography{combined}

\begin{thebibliography}{10}

\bibitem{Kachru:2003aw}
S.~Kachru, R.~Kallosh, A.~D. Linde, and S.~P. Trivedi, ``{De Sitter vacua in
  string theory},'' {\em Phys.Rev.} {\bf D68} (2003) 046005,
\href{http://www.arXiv.org/abs/hep-th/0301240}{{\tt hep-th/0301240}}.
%%CITATION = HEP-TH/0301240;%%.

\bibitem{Ooguri:1996me}
H.~Ooguri and C.~Vafa, ``{S}umming up {D}-instantons,'' {\em Phys. Rev. Lett.}
  {\bf 77} (1996) 3296--3298,
\href{http://www.arXiv.org/abs/hep-th/9608079}{{\tt hep-th/9608079}}.
%%CITATION = HEP-TH 9608079;%%.

\bibitem{Witten:1995ex}
E.~Witten, ``{String theory dynamics in various dimensions},'' {\em Nucl.
  Phys.} {\bf B443} (1995) 85--126,
\href{http://www.arXiv.org/abs/hep-th/9503124}{{\tt hep-th/9503124}}.
%%CITATION = HEP-TH/9503124;%%.

\bibitem{Green:1997di}
M.~B. Green and P.~Vanhove, ``{D instantons, strings and M theory},'' {\em
  Phys.Lett.} {\bf B408} (1997) 122--134,
\href{http://www.arXiv.org/abs/hep-th/9704145}{{\tt hep-th/9704145}}.
%%CITATION = HEP-TH/9704145;%%.

\bibitem{Bagger:1983tt}
J.~Bagger and E.~Witten, ``{M}atter couplings in {${\mathcal N}=2$}
  supergravity,'' {\em Nucl. Phys.} {\bf B222} (1983)
1.
%%CITATION = NUPHA,B222,1;%%.

\bibitem{deWit:1984px}
B.~de~Wit, P.~Lauwers, and A.~Van~Proeyen, ``{Lagrangians of N=2 Supergravity -
  Matter Systems},'' {\em Nucl. Phys.} {\bf B255} (1985)
569.
%%CITATION = NUPHA,B255,569;%%.

\bibitem{MR1327157}
C.~LeBrun, ``Fano manifolds, contact structures, and quaternionic geometry,''
  {\em Internat. J. Math.} {\bf 6} (1995), no.~3, 419--437,
\href{http://www.arXiv.org/abs/dg-ga/9409001}{{\tt dg-ga/9409001}}.
%%CITATION = DG-GA/9409001;%%.

\bibitem{Alexandrov:2008ds}
S.~Alexandrov, B.~Pioline, F.~Saueressig, and S.~Vandoren, ``{Linear
  perturbations of Hyperkahler metrics},'' {\em Lett. Math. Phys.} {\bf 87}
  (2009) 225--265,
\href{http://www.arXiv.org/abs/0806.4620}{{\tt 0806.4620}}.
%%CITATION = 0806.4620;%%.

\bibitem{Alexandrov:2008nk}
S.~Alexandrov, B.~Pioline, F.~Saueressig, and S.~Vandoren, ``{Linear
  perturbations of quaternionic metrics},'' {\em Commun. Math. Phys.} {\bf 296}
  (2010) 353--403,
\href{http://www.arXiv.org/abs/0810.1675}{{\tt 0810.1675}}.
%%CITATION = 0810.1675;%%.

\bibitem{RoblesLlana:2006is}
D.~Robles-Llana, M.~Ro\v{c}ek, F.~Saueressig, U.~Theis, and S.~Vandoren,
  ``{Nonperturbative corrections to 4D string theory effective actions from
  SL(2,Z) duality and supersymmetry},'' {\em Phys. Rev. Lett.} {\bf 98} (2007)
  211602,
\href{http://www.arXiv.org/abs/hep-th/0612027}{{\tt hep-th/0612027}}.
%%CITATION = HEP-TH/0612027;%%.

\bibitem{RoblesLlana:2007ae}
D.~Robles-Llana, F.~Saueressig, U.~Theis, and S.~Vandoren, ``{Membrane
  instantons from mirror symmetry},'' {\em Commun. Num. Theor. Phys.} {\bf 1}
  (2007) 681, \href{http://www.arXiv.org/abs/0707.0838}{{\tt 0707.0838}}.

\bibitem{Alexandrov:2008gh}
S.~Alexandrov, B.~Pioline, F.~Saueressig, and S.~Vandoren, ``{D-instantons and
  twistors},'' {\em JHEP} {\bf 03} (2009) 044,
\href{http://www.arXiv.org/abs/0812.4219}{{\tt 0812.4219}}.
%%CITATION = 0812.4219;%%.

\bibitem{Alexandrov:2009zh}
S.~Alexandrov, ``{D-instantons and twistors: some exact results},'' {\em J.
  Phys.} {\bf A42} (2009) 335402,
\href{http://www.arXiv.org/abs/0902.2761}{{\tt 0902.2761}}.
%%CITATION = 0902.2761;%%.

\bibitem{Alexandrov:2010ca}
S.~Alexandrov, D.~Persson, and B.~Pioline, ``{Fivebrane instantons, topological
  wave functions and hypermultiplet moduli spaces},'' {\em JHEP} {\bf 1103}
  (2011) 111, \href{http://www.arXiv.org/abs/1010.5792}{{\tt 1010.5792}}.

\bibitem{Alexandrov:2012au}
S.~Alexandrov, J.~Manschot, and B.~Pioline, ``{D3-instantons, Mock Theta Series
  and Twistors},'' {\em JHEP} {\bf 1304} (2013) 002,
\href{http://www.arXiv.org/abs/1207.1109}{{\tt 1207.1109}}.
%%CITATION = ARXIV:1207.1109;%%.

\bibitem{Alexandrov:2014mfa}
S.~Alexandrov and S.~Banerjee, ``{Fivebrane instantons in Calabi-Yau
  compactifications},'' {\em Phys.Rev.} {\bf D90} (2014) 041902,
\href{http://www.arXiv.org/abs/1403.1265}{{\tt 1403.1265}}.
%%CITATION = ARXIV:1403.1265;%%.

\bibitem{Alexandrov:2014rca}
S.~Alexandrov and S.~Banerjee, ``{Dualities and fivebrane instantons},'' {\em
  JHEP} {\bf 1411} (2014) 040,
\href{http://www.arXiv.org/abs/1405.0291}{{\tt 1405.0291}}.
%%CITATION = ARXIV:1405.0291;%%.

\bibitem{Alexandrov:2011va}
S.~Alexandrov, ``{Twistor Approach to String Compactifications: a Review},''
  {\em Phys. Rept.} {\bf 522} (2013) 1--57,
\href{http://www.arXiv.org/abs/1111.2892}{{\tt 1111.2892}}.
%%CITATION = ARXIV:1111.2892;%%.

\bibitem{Alexandrov:2013yva}
S.~Alexandrov, J.~Manschot, D.~Persson, and B.~Pioline, ``{Quantum
  hypermultiplet moduli spaces in N=2 string vacua: a review},''
\href{http://www.arXiv.org/abs/1304.0766}{{\tt 1304.0766}}.
%%CITATION = ARXIV:1304.0766;%%.

\bibitem{Polchinski:1995sm}
J.~Polchinski and A.~Strominger, ``{New vacua for type II string theory},''
  {\em Phys.Lett.} {\bf B388} (1996) 736--742,
\href{http://www.arXiv.org/abs/hep-th/9510227}{{\tt hep-th/9510227}}.
%%CITATION = HEP-TH/9510227;%%.

\bibitem{deWit:2001bk}
B.~de~Wit, M.~Ro\v{c}ek, and S.~Vandoren, ``Gauging isometries on hyperkaehler
  cones and quaternion- kaehler manifolds,'' {\em Phys. Lett.} {\bf B511}
  (2001) 302--310,
\href{http://www.arXiv.org/abs/hep-th/0104215}{{\tt hep-th/0104215}}.
%%CITATION = HEP-TH/0104215;%%.

\bibitem{MR1423177}
K.~P. Tod, ``The {${\rm SU}(\infty)$}-{T}oda field equation and special
  four-dimensional metrics,'' in {\em Geometry and physics ({A}arhus, 1995)},
  vol.~184 of {\em Lecture Notes in Pure and Appl. Math.}, pp.~307--312.
\newblock Dekker, New York, 1997.

\bibitem{Alexandrov:2009qq}
S.~Alexandrov and F.~Saueressig, ``{Quantum mirror symmetry and twistors},''
  {\em JHEP} {\bf 09} (2009) 108,
\href{http://www.arXiv.org/abs/0906.3743}{{\tt 0906.3743}}.
%%CITATION = 0906.3743;%%.

\bibitem{Antoniadis:2003sw}
I.~Antoniadis, R.~Minasian, S.~Theisen, and P.~Vanhove, ``String loop
  corrections to the universal hypermultiplet,'' {\em Class. Quant. Grav.} {\bf
  20} (2003) 5079--5102,
\href{http://www.arXiv.org/abs/hep-th/0307268}{{\tt hep-th/0307268}}.
%%CITATION = HEP-TH/0307268;%%.

\bibitem{Robles-Llana:2006ez}
D.~Robles-Llana, F.~Saueressig, and S.~Vandoren, ``String loop corrected
  hypermultiplet moduli spaces,'' {\em JHEP} {\bf 03} (2006) 081,
\href{http://www.arXiv.org/abs/hep-th/0602164}{{\tt hep-th/0602164}}.
%%CITATION = HEP-TH 0602164;%%.

\bibitem{Alexandrov:2007ec}
S.~Alexandrov, ``{Quantum covariant c-map},'' {\em JHEP} {\bf 05} (2007) 094,
\href{http://www.arXiv.org/abs/hep-th/0702203}{{\tt hep-th/0702203}}.
%%CITATION = HEP-TH/0702203;%%.

\bibitem{Becker:1995kb}
K.~Becker, M.~Becker, and A.~Strominger, ``Five-branes, membranes and
  nonperturbative string theory,'' {\em Nucl. Phys.} {\bf B456} (1995)
  130--152,
\href{http://www.arXiv.org/abs/hep-th/9507158}{{\tt hep-th/9507158}}.
%%CITATION = NUPHA,B456,130;%%.

\bibitem{Craps:1997gp}
B.~Craps, F.~Roose, W.~Troost, and A.~Van~Proeyen, ``{What is special Kahler
  geometry?},'' {\em Nucl.Phys.} {\bf B503} (1997) 565--613,
\href{http://www.arXiv.org/abs/hep-th/9703082}{{\tt hep-th/9703082}}.
%%CITATION = HEP-TH/9703082;%%.

\bibitem{Cecotti:1989qn}
S.~Cecotti, S.~Ferrara, and L.~Girardello, ``Geometry of type {I}{I}
  superstrings and the moduli of superconformal field theories,'' {\em Int. J.
  Mod. Phys.} {\bf A4} (1989)
2475.
%%CITATION = IMPAE,A4,2475;%%.

\bibitem{Ferrara:1989ik}
S.~Ferrara and S.~Sabharwal, ``{Q}uaternionic manifolds for type {II}
  superstring vacua of {C}alabi-{Y}au spaces,'' {\em Nucl. Phys.} {\bf B332}
  (1990)
317.
%%CITATION = NUPHA,B332,317;%%.

\bibitem{Antoniadis:1997eg}
I.~Antoniadis, S.~Ferrara, R.~Minasian, and K.~S. Narain, ``{$R^4$ couplings in
  M- and type II theories on Calabi-Yau spaces},'' {\em Nucl. Phys.} {\bf B507}
  (1997) 571--588,
\href{http://www.arXiv.org/abs/hep-th/9707013}{{\tt hep-th/9707013}}.
%%CITATION = HEP-TH/9707013;%%.

\bibitem{Gunther:1998sc}
H.~G{\"u}nther, C.~Herrmann, and J.~Louis, ``{Quantum corrections in the
  hypermultiplet moduli space},'' {\em Fortsch. Phys.} {\bf 48} (2000)
  119--123,
\href{http://www.arXiv.org/abs/hep-th/9901137}{{\tt hep-th/9901137}}.
%%CITATION = HEP-TH/9901137;%%.

\bibitem{Alexandrov:2010np}
S.~Alexandrov, D.~Persson, and B.~Pioline, ``{On the topology of the
  hypermultiplet moduli space in type II/CY string vacua},'' {\em Phys. Rev.}
  {\bf D83} (2011) 026001, \href{http://www.arXiv.org/abs/1009.3026}{{\tt
  1009.3026}}.

\bibitem{Sharpe:1999qz}
E.~R. Sharpe, ``{D-branes, derived categories, and Grothendieck groups},'' {\em
  Nucl. Phys.} {\bf B561} (1999) 433--450,
\href{http://www.arXiv.org/abs/hep-th/9902116}{{\tt hep-th/9902116}}.
%%CITATION = HEP-TH/9902116;%%.

\bibitem{Douglas:2000gi}
M.~R. Douglas, ``{D-branes, categories and N = 1 supersymmetry},'' {\em J.
  Math. Phys.} {\bf 42} (2001) 2818--2843,
\href{http://www.arXiv.org/abs/hep-th/0011017}{{\tt hep-th/0011017}}.
%%CITATION = HEP-TH/0011017;%%.

\bibitem{ks}
M.~Kontsevich and Y.~Soibelman, ``{Stability structures, motivic
  Donaldson-Thomas invariants and cluster transformations},''
  \href{http://www.arXiv.org/abs/0811.2435}{{\tt 0811.2435}}.

\bibitem{Zagier-dilog}
D.~Zagier, ``The dilogarithm function,'' in {\em Frontiers in Number Theory,
  Physics, and Geometry II}, pp.~3--65.
\newblock Berlin: Springer-Verlag, 2007.

\bibitem{Alexandrov:2011ac}
S.~Alexandrov, D.~Persson, and B.~Pioline, ``{Wall-crossing, Rogers
  dilogarithm, and the QK/HK correspondence},'' {\em JHEP} {\bf 1112} (2011)
  027,
\href{http://www.arXiv.org/abs/1110.0466}{{\tt 1110.0466}}.
%%CITATION = ARXIV:1110.0466;%%.

\bibitem{Alexandrov:2014wca}
S.~Alexandrov, G.~W. Moore, A.~Neitzke, and B.~Pioline, ``{An $R^3$ index for
  four-dimensional $N=2$ field theories},''
\href{http://www.arXiv.org/abs/1406.2360}{{\tt 1406.2360}}.
%%CITATION = ARXIV:1406.2360;%%.

\bibitem{Bohm:1999uk}
R.~B{\"o}hm, H.~G{\"u}nther, C.~Herrmann, and J.~Louis, ``{Compactification of
  type IIB string theory on Calabi-Yau threefolds},'' {\em Nucl. Phys.} {\bf
  B569} (2000) 229--246,
\href{http://www.arXiv.org/abs/hep-th/9908007}{{\tt hep-th/9908007}}.
%%CITATION = HEP-TH/9908007;%%.

\bibitem{Alexandrov:2012bu}
S.~Alexandrov and B.~Pioline, ``{S-duality in Twistor Space},'' {\em JHEP} {\bf
  1208} (2012) 112,
\href{http://www.arXiv.org/abs/1206.1341}{{\tt 1206.1341}}.
%%CITATION = ARXIV:1206.1341;%%.

\bibitem{Alexandrov:2013mha}
S.~Alexandrov and S.~Banerjee, ``{Modularity, Quaternion-Kahler spaces and
  Mirror Symmetry},'' {\em J. Math. Phys. 54,} {\bf 102301} (2013)
\href{http://www.arXiv.org/abs/1306.1837}{{\tt 1306.1837}}.
%%CITATION = ARXIV:1306.1837;%%.

\bibitem{Strominger:1997eb}
A.~Strominger, ``{Loop corrections to the universal hypermultiplet},'' {\em
  Phys. Lett.} {\bf B421} (1998) 139--148,
\href{http://www.arXiv.org/abs/hep-th/9706195}{{\tt hep-th/9706195}}.
%%CITATION = HEP-TH/9706195;%%.

\bibitem{Ketov:2001gq}
S.~V. Ketov, ``{Universal hypermultiplet metrics},'' {\em Nucl.Phys.} {\bf
  B604} (2001) 256--280,
\href{http://www.arXiv.org/abs/hep-th/0102099}{{\tt hep-th/0102099}}.
%%CITATION = HEP-TH/0102099;%%.

\bibitem{Ketov:2001ky}
S.~V. Ketov, ``{D instantons and universal hypermultiplet},''
\href{http://www.arXiv.org/abs/hep-th/0112012}{{\tt hep-th/0112012}}.
%%CITATION = HEP-TH/0112012;%%.

\bibitem{Ketov:2002vr}
S.~V. Ketov, ``{Summing up D-instantons in N = 2 supergravity},'' {\em Nucl.
  Phys.} {\bf B649} (2003) 365--388,
\href{http://www.arXiv.org/abs/hep-th/0209003}{{\tt hep-th/0209003}}.
%%CITATION = HEP-TH/0209003;%%.

\bibitem{Saueressig:2005es}
F.~Saueressig, U.~Theis, and S.~Vandoren, ``{On de Sitter vacua in type IIA
  orientifold compactifications},'' {\em Phys. Lett.} {\bf B633} (2006)
  125--128,
\href{http://www.arXiv.org/abs/hep-th/0506181}{{\tt hep-th/0506181}}.
%%CITATION = HEP-TH/0506181;%%.

\bibitem{Theis:2014bia}
U.~Theis, ``{Membrane Instantons from Toda Field Theory},''
\href{http://www.arXiv.org/abs/1408.4632}{{\tt 1408.4632}}.
%%CITATION = ARXIV:1408.4632;%%.

\bibitem{Alexandrov:2006hx}
S.~Alexandrov, F.~Saueressig, and S.~Vandoren, ``{Membrane and fivebrane
  instantons from quaternionic geometry},'' {\em JHEP} {\bf 09} (2006) 040,
\href{http://www.arXiv.org/abs/hep-th/0606259}{{\tt hep-th/0606259}}.
%%CITATION = HEP-TH/0606259;%%.

\bibitem{Przanowski:1984qq}
M.~Przanowski, ``{Locally Hermite Einstein, selfdual gravitational
  instantons},'' {\em Acta Phys. Polon.} {\bf B14} (1983)
625--627.
%%CITATION = APPOA,B14,625;%%.

\bibitem{Alexandrov:2009vj}
S.~Alexandrov, B.~Pioline, and S.~Vandoren, ``{Self-dual Einstein Spaces,
  Heavenly Metrics and Twistors},'' {\em J. Math. Phys.} {\bf 51} (2010)
  073510, \href{http://www.arXiv.org/abs/0912.3406}{{\tt 0912.3406}}.

\bibitem{Bao:2009fg}
L.~Bao, A.~Kleinschmidt, B.~E.~W. Nilsson, D.~Persson, and B.~Pioline,
  ``{Instanton Corrections to the Universal Hypermultiplet and Automorphic
  Forms on SU(2,1)},'' {\em Commun. Num. Theor. Phys.} {\bf 4} (2010) 187--266,
\href{http://www.arXiv.org/abs/0909.4299}{{\tt 0909.4299}}.
%%CITATION = 0909.4299;%%.

\bibitem{Candelas:1990rm}
P.~Candelas, X.~C. de~la Ossa, P.~S. Green, and L.~Parkes, ``{A pair of
  Calabi-Yau manifolds as an exactly soluble superconformal theory},'' {\em
  Nucl. Phys.} {\bf B359} (1991)
21--74.
%%CITATION = NUPHA,B359,21;%%.

\bibitem{Zamolodchikov:1989cf}
A.~B. Zamolodchikov, ``{Thermodynamic Bethe Ansatz in Relativistic Models.
  Scaling Three State Potts and Lee-Yang Models},'' {\em Nucl. Phys.} {\bf
  B342} (1990)
695--720.
%%CITATION = NUPHA,B342,695;%%.

\bibitem{Gaiotto:2008cd}
D.~Gaiotto, G.~W. Moore, and A.~Neitzke, ``{Four-dimensional wall-crossing via
  three-dimensional field theory},'' {\em Commun. Math. Phys.} {\bf 299} (2010)
  163--224, \href{http://www.arXiv.org/abs/0807.4723}{{\tt 0807.4723}}.

\bibitem{Alexandrov:2010pp}
S.~Alexandrov and P.~Roche, ``{TBA for non-perturbative moduli spaces},'' {\em
  JHEP} {\bf 1006} (2010) 066, \href{http://www.arXiv.org/abs/1003.3964}{{\tt
  1003.3964}}.

\bibitem{Pioline:2009ia}
B.~Pioline and S.~Vandoren, ``{Large D-instanton effects in string theory},''
  {\em JHEP} {\bf 07} (2009) 008,
\href{http://www.arXiv.org/abs/0904.2303}{{\tt 0904.2303}}.
%%CITATION = 0904.2303;%%.

\bibitem{Fre:2014pca}
P.~Fr\'e, A.~Sorin, and M.~Trigiante, ``{The $c$-map, Tits Satake subalgebras
  and the search for $\mathcal{N}=2$ inflaton potentials},''
\href{http://www.arXiv.org/abs/1407.6956}{{\tt 1407.6956}}.
%%CITATION = ARXIV:1407.6956;%%.

\bibitem{Ketov:2002fu}
S.~V. Ketov, ``{Instanton induced scalar potential for the universal
  hypermultiplet},'' {\em Nucl.Phys.} {\bf B656} (2003) 63--77,
\href{http://www.arXiv.org/abs/hep-th/0212003}{{\tt hep-th/0212003}}.
%%CITATION = HEP-TH/0212003;%%.

\bibitem{Ketov:2014roa}
S.~V. Ketov, ``{Natural Inflation and Universal Hypermultiplet},''
\href{http://www.arXiv.org/abs/1402.0627}{{\tt 1402.0627}}.
%%CITATION = ARXIV:1402.0627;%%.

\bibitem{deWit:1995jd}
B.~de~Wit and A.~Van~Proeyen, ``{Special geometry and symplectic
  transformations},'' {\em Nucl.Phys.Proc.Suppl.} {\bf 45BC} (1996) 196--206,
\href{http://www.arXiv.org/abs/hep-th/9510186}{{\tt hep-th/9510186}}.
%%CITATION = HEP-TH/9510186;%%.

\end{thebibliography}

\end{document}